\documentclass[english,12pt,aps,prd,a4paper,preprintnumbers,floatfix,nofootinbib,superscriptaddress, notitlepage]{revtex4-1} 
\usepackage[usenames,dvipsnames]{color}  
\usepackage{graphicx}
\usepackage{setspace}
\usepackage{caption}
\usepackage{subcaption}
\usepackage{amsmath}
\usepackage{amssymb}
\usepackage[colorlinks=true,citecolor=darkred,urlcolor=darkred, pdfborder={0 0 0}]{hyperref}
\definecolor{darkred}{rgb}{0.6,0,0}
\usepackage[normalem]{ulem}
\usepackage{xcolor}
\usepackage{array}
\usepackage{bm}
\usepackage{braket}

\makeatletter
\def\p@subsection{}
\makeatother

\definecolor{linkcolor}{rgb}{0,0,0.5}

\def\gsim{\raise0.3ex\hbox{$\;>$\kern-0.75em\raise-1.1ex\hbox{$\sim\;$}}}
\def\lsim{\raise0.3ex\hbox{$\;<$\kern-0.75em\raise-1.1ex\hbox{$\sim\;$}}}

\def\beqn#1{\begin{equation}\label{#1}}
\def\eeqn{\end{equation}}

\def\beqa#1{\begin{eqnarray}\label{#1}}
\def\eeqa{\end{eqnarray}}

\def\Z2{$\mathcal{Z_2}$}

\def\hc{\mathrm{h.c.}}

\newcommand {\ignore}[1]{}

\def\SM{$\mathrm{SU(3)_c \otimes SU(2)_L \otimes U(1)_Y}$ }

\def\BSM{$\mathrm{SU(3)_c \otimes SU(2)_L \otimes U(1)_Y \otimes \mathbb{Z}_2}$ }
 
\def\321{$\mathrm{SU(3) \otimes SU(2) \otimes U(1)}$ }

\newcommand{\AddrIISERB}{Department of Physics, Indian Institute of Science Education and Research - Bhopal, \\ 
Bhopal Bypass Road, Bhauri, Bhopal 462066, India}

\begin{document}

\title{\color{BrickRed}CDF-II $W$ Boson Mass Anomaly in the Canonical Scotogenic Neutrino-Dark Matter Model}
\author{Aditya Batra}\email{adityab17@iiserb.ac.in}
\affiliation{\AddrIISERB}
\author{ShivaSankar K.A.}\email{shivasankar17@iiserb.ac.in}
\affiliation{\AddrIISERB}
\author{Sanjoy Mandal}\email{smandal@kias.re.kr}
\affiliation{Korea Institute for Advanced Study, Seoul 02455, Korea}
\author{Hemant Prajapati}\email{hemant19@iiserb.ac.in}
\affiliation{\AddrIISERB}
\author{Rahul Srivastava}\email{rahul@iiserb.ac.in}
\affiliation{\AddrIISERB}

\begin{abstract}
\vspace{1cm} 

The CDF-II collaboration's recent high-precision measurement of $W$ boson mass indicates new physics contribution(s) beyond the Standard Model. We investigate the possibility of the well-known canonical Scotogenic model to explain the CDF-II measurement. The Scotogenic model is a popular scenario beyond the Standard Model that induces neutrino masses at the 1-loop level and includes a viable dark matter candidate, either scalar or fermionic. For both scalar and fermionic dark matter possibilities, we simultaneously examine the constraints coming from (a) neutrino mass, oscillation,  neutrinoless double beta decay and lepton flavour violation experiments, (b) from LEP and LHC (c) from dark matter relic density and direct detection experiments (d) from the oblique $S,T,U$ parameter values consistent with CDF-II $W$ boson measurement. We demonstrate that the new CDF-II measurement rules out the feasible parameter space of the scalar dark matter in the high mass regions ($m_{\eta_{R}} \gtrsim 500~\text{GeV}$), while still allowing the intermediate mass regions $54~\text{GeV} \lesssim m_{\eta_{R}} \lesssim 76~\text{GeV}$.  We also showed that the fermionic dark matter candidate in the canonical Scotogenic model, in the range $M_{N_{1}} \lesssim 500~\text{GeV} $ , can simultaneously explain all the aforementioned issues. Furthermore, we investigated how the recent findings from ATLAS 2023 impact this study.
\end{abstract}
\maketitle
\section{INTRODUCTION}\label{sec1-intro}
With the recent finding of the Higgs like boson with mass $125$ GeV at the LHC~\cite{Aad:2012tfa,Chatrchyan:2012xdj}, the Standard Model (SM) of particle physics has stood the test of time as a well-understood description of the world we observe at electroweak energy scales. But several experimental findings imply that SM cannot be an all-encompassing theory. The discovery of neutrino oscillations~\cite{Super-Kamiokande:1998kpq} and the existence of dark matter at cosmic scales~\cite{Planck:2018vyg} have provided irrefutable evidence for physics that the SM cannot explain. As a result, despite its many accomplishments, it is now widely expected that SM cannot be the final theory of nature. This leads us to investigate the extension of SM, which can account for aforementioned problems with SM. The CDF-II collaboration recently published their high precision measurement of the $W$ boson mass $m_W^{\rm CDF} = 80.4335 \pm 0.0094$ GeV~\cite{CDF:2022hxs}, which reveals a $7$-$\sigma$ difference from the SM expectation $m_W^{\rm SM} = 80.354 \pm 0.007 \text{GeV}$~\cite{Zyla:2020zbs}. Because of the large deviation from the SM prediction of $W$ mass, novel physics beyond Standard Model (BSM) is needed to explain the CDF-II measurement~\cite{Fan:2022dck,Lu:2022bgw,Athron:2022qpo,Yuan:2022cpw,Strumia:2022qkt,Yang:2022gvz,deBlas:2022hdk,Zhu:2022tpr,Du:2022pbp,Tang:2022pxh,Cacciapaglia:2022xih,Blennow:2022yfm,Sakurai:2022hwh,Arias-Aragon:2022ats,Fan:2022yly,Liu:2022jdq,Lee:2022nqz,Cheng:2022jyi,Bagnaschi:2022whn,Paul:2022dds,Bahl:2022xzi,Asadi:2022xiy,DiLuzio:2022xns,Athron:2022isz,Gu:2022htv,Heckman:2022the,Babu:2022pdn,Zhu:2022scj,Balkin:2022glu,Biekotter:2022abc,Endo:2022kiw,Crivellin:2022fdf,Cheung:2022zsb,Du:2022brr,Heo:2022dey,Krasnikov:2022xsi,Ahn:2022xeq,Han:2022juu,Zheng:2022irz,Perez:2022uil,Ghoshal:2022vzo,Kawamura:2022uft,Nagao:2022oin,Kanemura:2022ahw,Mondal:2022xdy,Zhang:2022nnh,Borah:2022obi,Chowdhury:2022moc,Arcadi:2022dmt,Cirigliano:2022qdm,Bagnaschi:2022qhb,Popov:2022ldh,Bhaskar:2022vgk,Du:2022fqv,Ghorbani:2022vtv,Carpenter:2022oyg,Almeida:2022lcs,Cheng:2022aau,Addazi:2022fbj,Cao:2022mif,Heeck:2022fvl,Baek:2022agi,Borah:2022zim,Lee:2022gyf,Ahn:2022xeq} which we aim to explore in this work.

There has been significant interest in recent times in models that potentially link various shortcoming of the SM. In this paper, we analyze the canonical Scotogenic model first proposed by Ma~\cite{Ma:2006km}. It is one of the simplest models which can explain both neutrino mass generation along with a viable dark matter candidate particle. As we discuss in details in Sec.~\ref{sec:scoto-model}, the canonical Scotogenic model is extremely simple and elegant. It adds just two type of particles to the particle content of the SM: an $SU(2)_L$ scalar doublet $\eta$ and $SU(2)_L$ singlet fermions $N$. In addition a dark parity symmetry $\mathbb{Z}_2$ is also added to the symmetries of the SM. The dark $\mathbb{Z}_2$ symmetry ensures the stability of the dark matter particle in the model. The model then generates neutrino mass at one loop with the $\eta$ and $N$ running in the loop. More details of the model are discussed in Sec.~\ref{sec:scoto-model}. 

In this paper we show that despite its simplicity, the canonical Scotogenic model has enough flexibility to account for the CDF-II anomaly in mass of $W$ boson measurement. This is achieved through loop corrections to gauge boson two-point functions which can be quantified through the oblique parameter $S$, $T$, $U$. The scotogenic model can account for the CDF II W boson mass measurements through the oblique parameter corrections. We further show that the model has a parameter space simultaneously consistent with the best fit values of $S$, $T$, $U$ parameters with the CDF-II results along with the constraints coming from neutrino and dark matter physics.

The plan of the paper is as follows. In Section~\ref{sec:scoto-model}, we summarize the Scotogenic model obtaining the scalar mass spectrum, pertubativity and stability constraints. We also discuss the one loop Sotogenic neutrino mass generation and the constraints coming from KATRIN and cosmological neutrino mass limits, neutrino oscillations and neutrinoless double beta decay experiments. In Section~\ref{sec:stu}, we discuss the $S$, $T$, and $U$ parameter space for the model considering new CDF-II results along with the neutrino physics constraints as well as constraints coming from lepton flavour violating processes. In Section~\ref{sec:drak-matter}, we discuss the dark matter constraints for both the  scalar and fermionic dark matter possibilities taking into account the $S$, $T$, and $U$, neutrino physics and lepton flavour violating constraints.  Finally, we summarize our results in Section~\ref{sec:conclusion}.
\section{The canonical SCOTOGENIC MODEL}
\label{sec:scoto-model}
In this section, we discuss in details the canonical Scotogenic model~\cite{Ma:2006km} which is a simple extension of SM intimately linking the neutrino mass generation with WIMP dark matter. Since the initial proposal, the scotogenic mechanism has proved to be immensely popular and many different extensions of the basic mechanism have been proposed over the years, for some recent works see Refs.~\cite{Rojas:2018wym,Bonilla:2018ynb,Avila:2019hhv,CentellesChulia:2019gic,Kang:2019sab,Srivastava:2019xhh,Leite:2019grf,Barreiros:2020gxu,Beniwal:2020hjc,Mandal:2021yph,Borah:2021rbx,Escribano:2021wud,Sarazin:2021nwo,deBoer:2021pon}. However, the original Scotogenic model of Ma~\cite{Ma:2006km} still stands out for its simplicity and elegance. In the case of Scotogenic model, the SM symmetries are augmented with a new dark $\mathbb{Z}_2$ parity symmetry which should remain unbroken. Furthermore, the SM particle content is extended by just two types of ``dark sector'' particles: new $SU(2)_L$ doublet scalar $\eta$ and singlet fermions $N$ both of which are odd, while all SM particle are even under the $\mathbb{Z}_2$ symmetry. Owing to the presence of dark $\mathbb{Z}_2$ symmetry, the lightest dark sector particle in the model is stable and will be a good candidate for dark matter. The lepton and scalar particle content and the charge assignment of the particles are given in Tab.\ref{tab:tab1}.
\begin{table}[h]
	\begin{center}\label{tab:tab1}
		\begin{tabular}{| c| c || c | c |}
			\hline
			\hspace{0.5cm}Fields \hspace{0.5cm}   & \hspace{0.5cm} \SM \hspace{0.5cm} & \hspace{0.5cm} $\mathbb{Z}_2$ \hspace{0.5cm} \\
			\hline \hline
			$L$ \ \             & $(1, 2, -\frac{1}{2})$\ \      & $+1$ \\
			$ \ell$ \ \        & $(1, 1, -1)$\ \                    & $+1$ \\
			$\Phi$       \ \  & $(1, 2, \frac{1}{2})$\ \        & $+1$  \\
			\hline \hline 
			$ N$ \ \           & $(1,1,0)$ \ \ 					    &$-1$ \\
			$\eta$			\ \ &$(1,2, \frac{1}{2})$ 				  \ \			&$-1$ \\
			\hline \hline
		\end{tabular}
	\end{center}
	\caption{Particle content and  charge assignments for the canonical Scotogenic model.  The flavour indices are suppressed for brevity. $N$ and $\eta$ being odd under the $\mathbb{Z}_2$ dark parity, belong to the ``dark sector'' of the model.}
	\label{tab:tab1}
\end{table} 

The \BSM invariant Yukawa Lagrangian relevant for neutrino mass generation is given as
\begin{equation}\label{eq:yukawa}
	- \mathcal{L}_Y =  Y^N_{\alpha\beta} \overline{N}_{\alpha} \eta L_{\beta}  +  \frac{1}{2} \overline{N^c}_{\alpha} M_{\alpha \beta} N_{\beta}  +\hc ,
\end{equation}
where $Y^N_{\alpha\beta}$ is the Yukawa matrix and  $M_{\alpha\beta}$ is the Majorana mass matrix of the singlet fermions $N$. The current neutrino oscillation data implies that at least two neutrinos should be massive, which in turn means that we should have at least two generations of the fermion $N$. There is no maximum limit on the number of generations of $N$, however three generations are enough to generate masses for all three neutrinos. Here, we will use the minimum required two generations of $N$ for our analysis, however our main conclusions do not change even if more generations of $N$ are added to the model.
 
The scalar potential of the model is given by 
\begin{align}
	V &= m^2_{\Phi} \Phi^{\dagger} \Phi +  m_{\eta}^2 \eta^{\dagger} \eta + \frac{\lambda_1}{2} (\Phi^{\dagger} \Phi)^2 + \frac{\lambda_2}{2} (\eta^{\dagger}\eta)^2 + \lambda_3 (\Phi^{\dagger}\Phi)(\eta^{\dagger}\eta) \nonumber
	\\
	&+ \lambda_4 (\Phi^{\dagger} \eta)(\eta^{\dagger} \Phi) + \frac{\lambda_5}{2} \left[ (\Phi^{\dagger}\eta)^2 + (\eta^{\dagger}\Phi)^2 \right].
\end{align}
The requirement to have a stable minimum for the potential implies following conditions
\begin{align}\label{eq:vacuum}
	\lambda_1,\,\, \lambda_2 \geq 0 ; \hspace{1.5cm} \lambda_3,\,\, \lambda_3 + \lambda_4 - \vert \lambda_5 \vert > -2 \sqrt{\lambda_1 \lambda_2}.
\end{align}
The components of the $SU(2)_L$ doublet scalars are given by 
\begin{align}
	\Phi =&
	\begin{pmatrix}
		\phi^+ \\
		\phi^0 \\
	\end{pmatrix},
	\hspace{1cm}
	\eta = 
		\begin{pmatrix}
		\eta^+ \\
		\eta^0 \\
	\end{pmatrix}. 
	\hspace{2cm}
\end{align}
The electroweak symmetry is broken by the vacuum expectation value (VEV) of $\Phi$. In order to ensure the stability of the dark matter in the model, the dark  $\mathbb{Z}_2$ symmetry should remain unbroken. This in turn means that the field $\eta$ should not acquire any VEV. Thus, after electroweak symmetry breaking, we have
\begin{align}
	\langle \Phi \rangle = \frac{v_{\Phi}}{\sqrt{2}}, \hspace{1cm} \langle \eta \rangle = 0.
\end{align}
The presence of two $SU(2)_L$ doublets in the model means that after electroweak symmetry breaking there are five physical scalars; two CP even scalars $h,\eta_R$, one CP odd scalar $\eta_I$ and the electrically charged scalars $\eta^{\pm}$. The mass relations for the Higgs and the charged scalar fields are now given by
\begin{align}
	m_{h}^2 &= \lambda_1 v_{\Phi}^2,
	\\
	m_{\eta^{\pm}}^2 &= m_{\eta}^2 + \frac{\lambda_3}{2} v_{\Phi}^2 \label{eq:c-eta-m}.
\end{align}
The CP even scalar $h$ can be identified with the $125$ GeV particle discovered at LHC~\cite{Aad:2012tfa,Chatrchyan:2012xdj}. Masses of the dark sector neutral scalars $\eta_{R}$ and $\eta_{I}$ are given by 
\begin{align}
	m_{\eta_{R}}^2 &=  m_{\eta}^2 + (\lambda_3 + \lambda_4 + \lambda_5) \frac{v_{\Phi}^2}{2},
	\\
	m_{\eta_{I}}^2 &= m_{\eta}^2 + (\lambda_3 + \lambda_4 - \lambda_5) \frac{v_{\Phi}^2}{2}.
\end{align}
Notice that the $\lambda_5$ coupling leads to a mass splitting between the CP odd and even components of $\eta$ and the sign of $\lambda_5$ will determine which of the two will be lighter. For later convenience, we also define the parameter $m_{\eta^0}$ and $\lambda_{345}$ given by
\begin{equation}\label{eq:0-eta-m}
	m_{\eta^0}^2 = \frac{m_{\eta_R}^2 + m_{\eta_I}^2}{2},~~~\lambda_{345}= \lambda_{3}+\lambda_{4}+\lambda_{5}.
\end{equation}

As we have mentioned before, the lightest dark sector particle in the model will be the dark matter candidate. Since both $\eta$ and $N$ belong to the dark sector, the model has two dark matter possibilities, namely a scalar dark matter and a fermionic dark matter. In this work, we will examine both these possibilities.
For the scalar dark matter case, we again have two options; $\eta_R$ being the lightest dark sector particle or $\eta_I$ being the lightest dark sector particle. In this work to avoid repeating similar analysis twice, we will assume that $\eta_R$ is lighter than $\eta_I$. It should be noted that taking  $\eta_I$ lighter than $\eta_R$ leads to very similar results leaving all our conclusions unchanged.
The demand that the neutral scalar $\eta_R$ is the lightest dark scalar particle, imposes the following  additional constraints on the couplings given by 
\begin{equation}
	\lambda_4 + \lambda_5 <0; \hspace{2cm} \lambda_5 < 0 .
\end{equation}
Note here that the first condition ensures $m_{\eta_R} < m_{\eta^{\pm}}$ while the second ensures that $m_{\eta_R} < m_{\eta_I}$. Needless to say that if $\eta_R$ is the dark matter candidate then it should also be lighter than the dark fermions i.e. $m_{\eta_R} < M_{N}$.

\subsection{Neutrino masses}
\label{subsec:neut-masses}
\begin{figure}[!htbp]
	\centering
	\includegraphics{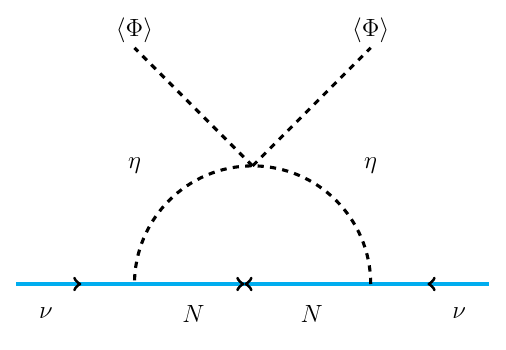}
	\caption{One loop neutrino mass in the minimal Scotogenic model, where $\eta^0 = (\eta_R, \eta_I)$}
	\label{fig:scotoloop}
\end{figure}

The light neutrino masses are generated at one loop in the Scotogenic model as shown in Fig.\ref{fig:scotoloop}. The odd particles under the $\mathbb{Z}_2$ symmetry participate in the loop. As we will discuss in Section~\ref{sec:stu}, the doublet scalar $\eta$ responsible for the neutrino mass generation plays a crucial role in explaining $W$ mass. Thus in the canonical Scotogenic model, the existence of dark matter, small neutrino masses and $W$ boson anomaly have a common origin. 

After electroweak symmetry breaking, the neutrino masses arising from the loop can be written as,  

\begin{align}\label{eq:neutrinomass}
	\mathcal{M}_{\alpha \beta}^{\nu} = \sum_i \frac{Y^N_{\alpha i} Y^N_{\beta i}}{32 \pi^2} M_{N_i} \left[ \frac{m_{\eta_R}^2}{m_{\eta_R}^2 - M_{N_i}^2} \log \left( \frac{m_{\eta_R}^2}{M_{N_i}^2} \right)  
	-  \frac{m_{\eta_I}^2}{m_{\eta_I}^2 - M_{N_i}^2} \log \left( \frac{m_{\eta_I}^2}{M_{N_i}^2} \right) \right],	
\end{align}
where $\alpha = 1,2,3$ are the three generations of lepton doublets and for the case of minimal Scotogenic model $i =1,2$ represent the two copies of the dark fermion $N$. The neutrino mass matrix $\mathcal{M}_{\alpha\beta}^{\nu}$ can also be expressed as 
\begin{equation}
	\mathcal{M}_{\alpha \beta}^{\nu} = \Big((Y^N)^T \Lambda Y^N \Big)_{\alpha \beta},\,\,\text{with}\,\, \Lambda=\begin{pmatrix}
		\Lambda_1 & 0  \\
		0& \Lambda_2 \\
	\end{pmatrix},
\end{equation}
where $\Lambda$ is the loop function given by 
\begin{align}
	 \Lambda_i = \frac{1}{32 \pi^2} M_{N_i} \left[ \frac{m_{\eta_R}^2}{m_{\eta_R}^2 - M_{N_i}^2} \log \left( \frac{m_{\eta_R}^2}{M_{N_i}^2} \right) 
	-  \frac{m_{\eta_I}^2}{m_{\eta_I}^2 - M_{N_i}^2} \log \left( \frac{m_{\eta_I}^2}{M_{N_i}^2} \right) \right].	
\end{align}
Using the Casas-Ibarra parametrization, the Yukawa terms can be written as~\cite{Casas:2001sr}
\begin{equation}\label{eq:casasibarra}
	Y^N = \sqrt{\Lambda^{-1}} R \sqrt{m_{\nu}} U_{\text{lep}}^{\dagger}
\end{equation}
where $U_{\text{lep}}$ is the leptonic mixing matrix, $m_{\nu}$ is the diagonal matrix of neutrino mass eigenvalues and $R$ is a complex rotation matrix. 

Since the minimal model has only two copies of the fermion $N$, there is an interesting consequence. In this case one of the light neutrino mass eigenstates will be zero and as a result there will be a lower bound on the amplitude $\braket{m_{\beta\beta}}$ of neutrinoless double beta decay~($0\nu\beta\beta$)~\cite{Reig:2018ztc,Barreiros:2018bju,Mandal:2019ndp}. For heavy TeV-scale neutrino mediators $N$, the main contribution to $\braket{m_{\beta\beta}}$ comes from the light active neutrinos. Using the symmetrical parametrization of the lepton mixing matrix $U_{\text{lep}}$~\cite{Schechter:1980gr}, it can be expressed as~\cite{Rodejohann:2011vc} 
%
\begin{align}
& \braket{m_{\beta\beta}}\approx \Big|\sum_i U_{\text{lep},ei}^2 \,m_i\Big|\nonumber\\
&=\Big|\cos^2\theta_{12}\cos^2\theta_{13}m_1+\sin^2\theta_{12}\cos^2\theta_{13}m_2 e^{2i\phi_{12}}+\sin^2\theta_{13}m_3 e^{2i\phi_{13}}\Big|.
\end{align} 
where $m_i$ and $\theta_{ij}$ are the light neutrino mass eigenstates and neutrino oscillation parameters. In our case, either $m_1=0$ or $m_3=0$ for normal and inverted neutrino mass ordering, respectively. There is effectively only one Majorana phase~(NO: $\phi\equiv \phi_{12}-\phi_{13}$, IO: $\phi\equiv \phi_{12}$).
\begin{figure}[htb!]
\includegraphics[width=0.95\textwidth]{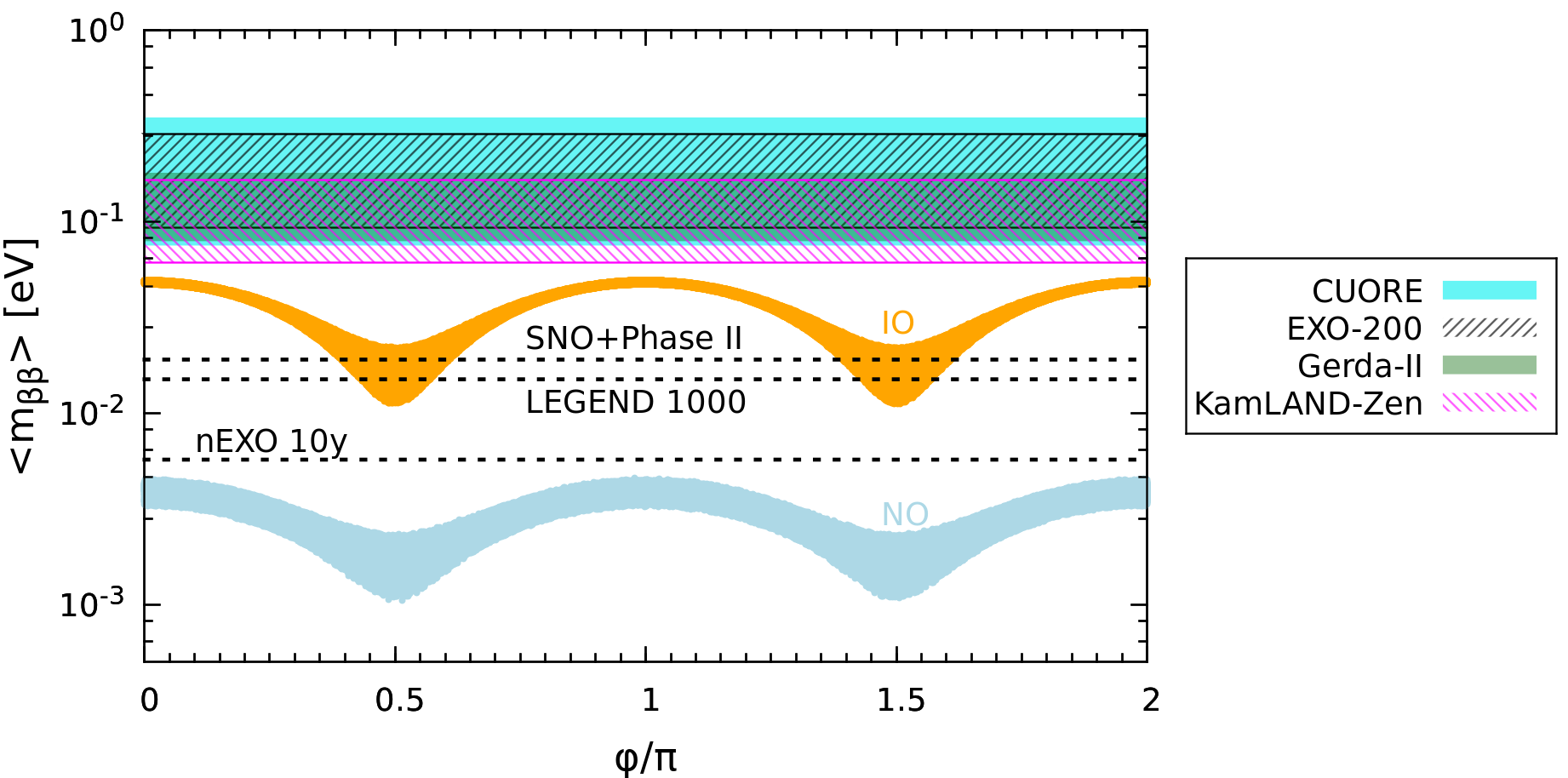}
\caption{
  Amplitude for $0\nu\beta\beta$ as a function of the Majorana phase when one neutrino is massless.  
  The light-blue and light-orange bands are the 3-$\sigma$ C.L. regions allowed by current oscillation experiments for normal and inverted mass ordering~\cite{deSalas:2020pgw}.
  The horizontal bands are the current limits from various $0\nu\beta\beta$ experiments, while the black lines show projected sensitivities. }
\label{fig:0nubb}
\end{figure}
\par In Fig.~\ref{fig:0nubb} we show the $\braket{m_{\beta\beta}}$ dependence on the Majorana phase $\phi$. The lower blue band is allowed for normal ordering, whereas the orange band corresponds to the inverted ordering. The bands are due to the allowed spread in neutrino oscillation parameters~\cite{deSalas:2020pgw}. Note that when there are three copies of $N$, there can be a cancellation amongst the light neutrino amplitudes for normal ordering, so that $\braket{m_{\beta\beta}} \to 0$. However, such a cancellation cannot happen when one neutrino is massless leading to a lower theoretical bound on $\braket{m_{\beta\beta}}$ even for normal ordering. The different bands in Fig.~\ref{fig:0nubb} show the current bounds from various experiments such as: CUORE~(cyan, $\braket{m_{\beta\beta}}<0.075-0.350$~eV)~\cite{CUORE:2020ymk}, EXO-200~(gray, $\braket{m_{\beta\beta}}< 0.093-0.286$~eV)~\cite{EXO-200:2019rkq}, GERDA-II~(green, $\braket{m_{\beta\beta}}<0.079-0.180$~eV)~\cite{GERDA:2020xhi} and KamLAND-Zen~(magenta, $\braket{m_{\beta\beta}}<0.061-0.165$~eV)~\cite{KamLAND-Zen:2016pfg}, whereas the three horizontal dashed black lines are the projected sensitivities of upcoming experiments: SNO+ Phase-II (0.019 eV)~\cite{SNO:2015wyx}, LEGEND-1000 (0.015 eV)~\cite{LEGEND:2017cdu} and nEXO - 10yr (0.0057 eV)~\cite{nEXO:2017nam}. Hence, one can expect that upcoming experiments will be able to probe the $\braket{m_{\beta\beta}}$ and the relevant Majorana phase, at least for inverted ordering.

The minimal Scotogenic model provides two types of dark matter candidates, the scalar candidate corresponding to the neutral component of $\eta$ ($\eta^R$ or $\eta^I$)\footnote{Since the results for $\eta^I$ dark matter are very similar to the case of $\eta^R$ dark matter, to avoid repetition, in this work we always take $\eta^R$ to be dark matter. } and the fermionic candidate corresponding to the lightest Majorana fermion $N_1$. We discuss both possibilities in the following sections in detail. Our aim is to check whether we can explain both the dark matter and CDF-II $W$ mass measurements in a consistent way.
\section{The $W$ Boson mass corrections from dark sector}
\label{sec:stu}
In the canonical Scotogenic model, the presence of the dark scalar $\eta$ leads to corrections to gauge boson two-point functions through loop diagrams as shown in Fig.~\ref{fig:selfenergy}.
The primary BSM effects can be parametrized by three gauge boson self-energy parameters dubbed the oblique parameters $S$, $T$ and $U$. These are the reparametrizations of the variables $\Delta\rho$, $\Delta\kappa$, and $\Delta r$, respectively, which absorb the radiative corrections to the overall $Z$ coupling strength, the effective weak mixing angle, and the $W$ mass. 
The corrections to the $W$ boson mass in terms of the  $S$, $T$, $U$ are given by 
\begin{align}
	m_W =m_W^{\rm SM}\Big[1-\frac{\alpha_{\rm em}}{4(c_W^2-s_W^2)}(S-1.55T-1.24 U)\Big]. 
\end{align}
Ref.~\cite{Lu:2022bgw} recently provided the values of these parameters based on a study of precision electroweak data, including the new CDF-II $W$-mass result to be:
\begin{align}\label{eq:STUnonzero}
	S=0.06\pm 0.10,\,\,  T=0.11\pm 0.12 \,\,\,\,\text{and}\,\,\,\, U=0.14\pm 0.09,
\end{align}
at 1-$\sigma$ with the correlation 
\begin{align}\label{eqcorrel}
	\rho_{ST}=0.90,\,\,  \rho_{SU}=-0.59 \,\,\,\,\text{and}\,\,\,\, \rho_{TU}=-0.85.
\end{align}
On the other hand, since the values of the $U$ parameter are found to be very small in many new physics models, it is reasonable to consider $U\approx 0$. With this assumption, Ref.~\cite{Lu:2022bgw} found the following values of $S$ and $T$:
\begin{figure}
	\includegraphics{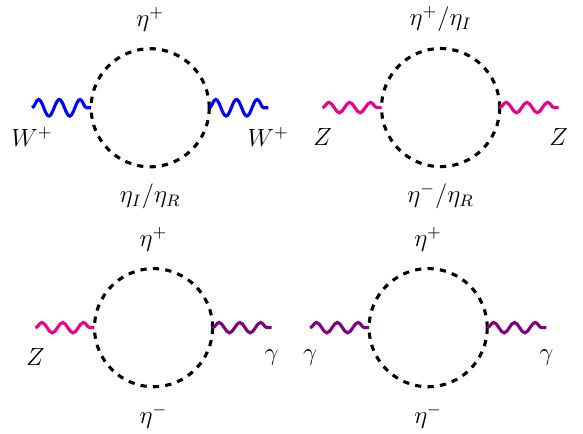}
	\caption{One loop vacuum polarization diagrams that contribute to the oblique parameters $S$, $T$, $U$.}
	\label{fig:selfenergy}
\end{figure}
\begin{align}
	S=0.14\pm 0.08 \,\,\,\text{and}\,\,\, T=0.26\pm 0.06\,\,\text{with the correlation}\,\, \rho_{ST}=0.93 .
\end{align}

In the canonical Scotogenic model, the dominant corrections to the oblique parameters comes from the dark scalar $\eta$ which is a doublet under $SU(2)_L$ symmetry. The scalar $\eta$ can contribute to the $W^{\pm},Z$ and $ \gamma$ vacuum polarization diagrams as shown in Figure~\ref{fig:selfenergy}. Explicit forms of $S$, $T$ and $U$ is derived from \cite{Branco:2011iw}. We have written the Explicit form in Apendix:\eqref{APA}.

\begin{table}[!htbp]
	\begin{center}
		\begin{tabular}{ | >{\centering\arraybackslash}m{1in} |>{\centering\arraybackslash}m{3in}| }
			\hline
			Parameter & Range \\
			\hline
			&\\[-12pt]
			$\lambda_2$  & $\bm{[} \, 10^{-6},  \sqrt{4 \pi} \, \bm{]} \; $ \\
			$\lambda_3$  & $\bm{[} \,  10^{-6},  \sqrt{4 \pi}  \, \bm{]} \; $ \\ 
			$\lambda_4$  & $\bm{[} \,-\sqrt{4 \pi},  -10^{-6} \, \bm{]} \; $ \\
			$\lambda_5$  & $\bm{[} \, -\sqrt{4 \pi},  -10^{-6}\, \bm{]} \; $ \\
			$\lambda_{345}$  & $\bm{[} \, -3,  3\, \bm{]} \; $ \\			
			$M_{N_1}$  & $\bm{[} \, m_{\eta^0} + \delta, 1000\; m_{\eta^0} \, \bm{]} \text{GeV} $ \\
			$M_{N_2}$  & $\bm{[} \, M_{N_1} + \delta, 1000\; m_{\eta^0} \, \bm{]} \text{GeV} $ \\
			$m_{\eta}^2$  & $\bm{[} \, 1, 10^8 \, \bm{]} \text{GeV}$ \\ 
			\hline
		\end{tabular}
	\end{center}
	\caption{Value range for the numerical parameter scan for $S$, $T$ and $U$ parameters and scalar dark matter analysis. Here, $\delta$ is small positive number and we maintained a hierarchy $m_{\eta_{0}} < M_{N_{1}} < M_{N_{2}}$ }
	\label{tab:scalar-dark-matter}
\end{table}

\begin{figure}[!htbp]
	\includegraphics[scale=0.3]{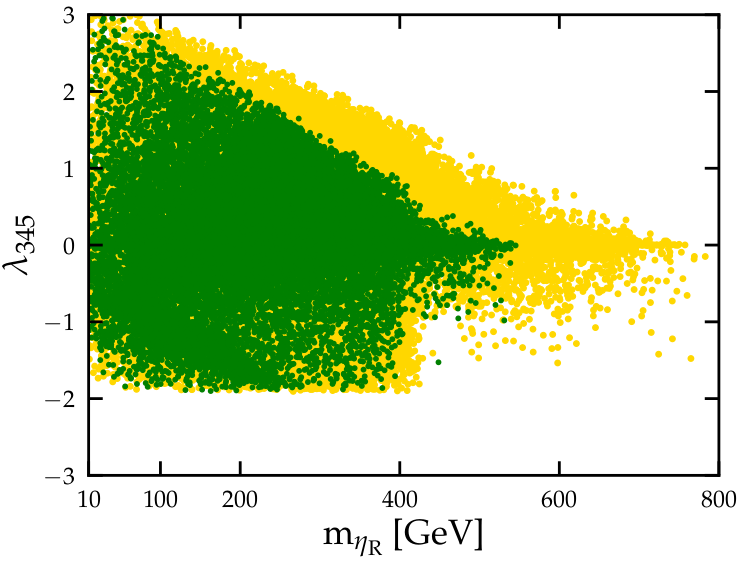}
	\includegraphics[scale=0.3]{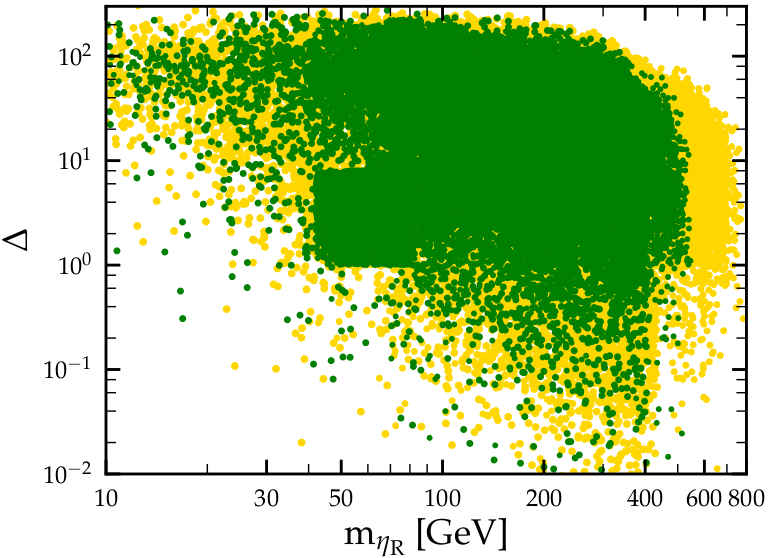}
	\includegraphics[scale=0.3]{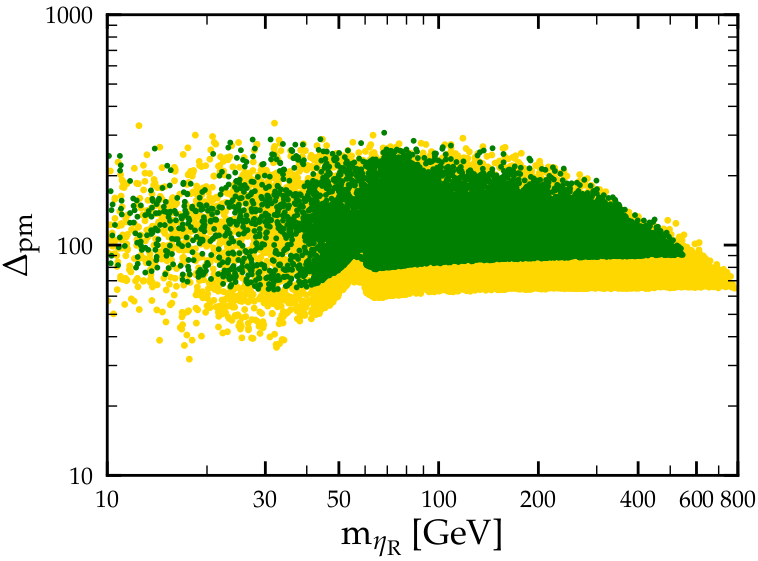}
	\caption{Correlation between revised values of the oblique  parameters after CDF II measurements and other model parameters. The yellow points satisfy revised $S$, $T$, and $U$ values as mentioned in Eq.~\ref{eq:STUnonzero}, at $3\sigma$, but not CDF II, W boson mass measurements. The green points satisfy $S$, $T$, and $U$ values and also CDF II, W mass at $1\sigma$. $\Delta$ is the mass difference of inert scalars ($m_{\eta_{I}} - m_{\eta_{R}} $), whereas $\Delta_{\text{pm}}$ is the mass difference between charged scalar and $\eta_{R}$ ($m_{\eta^{+}} - m_{\eta_{R}} $)}
	\label{fig:STUV}
\end{figure}

Fig.~\ref{fig:STUV} shows the correlations of various model parameters with oblique parameters, with $\eta_{R}$ being the lightest particle in the dark sector. Our numerical scan was performed varying the input parameters as given in Tab.~\ref{tab:scalar-dark-matter}. All the points in this plot satisfy the following constraints:
\begin{itemize}
\item We ensure that the scalar potential is bounded from below, by imposing the vacuum stability constraints given in Eq.~\ref{eq:vacuum}.
\item We ensure the perturbativity of the couplings, i.e. the scalar quartic couplings are taken to be less than $\mathcal{O}(1)$.
\item One of the key goals of the Scotogenic model is to provide an explanation for small neutrino masses. We guarantee this by demanding compatibility of allowed parameter range with the best-fit ranges of the neutrino oscillation parameters throughout our analysis. We impose this by using Eq.~\ref{eq:casasibarra}. For simplicity, the yet unknown Dirac and Majorana phases are set to zero and we have assumed normal ordering of the light neutrino masses.
\item The constraints from neutrinoless double beta decay (for normal ordering) experiments~\cite{CUORE:2020ymk,EXO-200:2019rkq,GERDA:2020xhi,KamLAND-Zen:2016pfg} as well as limits from cosmology~\cite{Planck:2018vyg} and KATRIN~\cite{KATRIN:2021uub} experiments are imposed.
\item The lepton flavour violating (LFV) processes $\ell_\alpha\to\ell_\beta\gamma$
impose  a very strong constraint on the model. The branching ratio of the process is computed as~\cite{Toma:2013zsa, Lindner:2016bgg}
\begin{equation}
\mathrm{Br}\left(\ell_\alpha\to\ell_\beta\gamma\right)=
\frac{3\alpha_\mathrm{em}}{64\pi^2 G_F^2 m_{\eta^+}^4}\left|
\sum_{i=1}^{2} Y^N_{i\alpha}(Y^N_{i\beta})^*F_2\left(\frac{M_{N_i}^2}{m_{\eta^+}^2}\right)
\right|^2
\mathrm{Br}\left(\ell_\alpha\to\ell_\beta\nu_\alpha\overline{\nu_{\beta}}\right)\ ,
\label{eq:lfv}
\end{equation}
where $F_2(x)$ is the loop function given in Ref.~\cite{Toma:2013zsa}.  The current experimental upper bounds for these processes~\cite{MEG:2013oxv, ParticleDataGroup:2016lqr, MEG:2016leq} are,
\begin{eqnarray}
\mathrm{Br}(\mu\to e\gamma) & \leq & 4.2\times10^{-13}\ ,\\
\mathrm{Br}(\tau\to \mu\gamma) &\leq& 4.4\times10^{-8}\ ,\\
\mathrm{Br}(\tau\to e\gamma) &\leq& 3.3\times10^{-8}.
\end{eqnarray}
In addition to these LFV constraints, we also impose the perturbativity constraints: $\text{Trace}(Y_N^\dagger Y_N)<4\pi$.
\end{itemize}
The yellow points in Fig.~\ref{fig:STUV} satisfy revised $S$, $T$, and $U$ values as mentioned in Eq.~\ref{eq:STUnonzero}, at $3\sigma$, but not CDF II, W boson mass measurements. The green points satisfy $S$, $T$, and $U$ values and also CDF II, W mass at $1\sigma$.
The updated $S$, $T$ and $U$ can be satisfied at $m_{\eta_{R}} \lesssim 800~\text{GeV} $, while the CDF II mass measurement can be satisfied at $m_{\eta_{R}} \lesssim 500~\text{GeV} $, as shown in Fig.~\ref{fig:STUV}. Note that large quartic couplings are necessary in the region with high dark matter mass in order to meet $S$, $T$, and $U$. Hence $m_{\eta_{R}} \gtrsim 500~ \text{GeV} $ is ruled out by the revised $S$, $T$, and $U$ values after CDF II results. As the $T$ parameter dominates the $W$ mass correction, a significant mass split between the $\eta^{+}$ and $\eta_{R}$ bosons can significantly increase the $W$-boson mass. The range of $\Delta_{\rm pm}~(\equiv m_{\eta^\pm} - m_{\eta_R})$ to satisfy the CDF II anomaly is $30~\text{GeV} \lesssim \Delta_{\rm pm} \lesssim 300~\text{GeV}$, while range of $\Delta(\equiv m_{\eta_I} - m_{\eta_R})$ is $0.01~\text{GeV} \lesssim \Delta \lesssim 300~\text{GeV}$.
We also add additional LEP and LHC constraints to the previously listed constraints in order to scan the whole parameter space:
\begin{itemize}	
\item LEP imposes very strong constraints on this model. We ensure that all the LEP constraints are included in our analysis \cite{Belyaev:2016lok}.

\begin{itemize}
\item $W$ and $Z$ boson widths are precisely measured at LEP. In this model, the scalar doublet $\eta$, couples to $W$ and $Z$ through the gauge coupling and is independent of any other model parameters. Hence we need to kinematically forbid the decay channel of $Z$ ($Z \rightarrow \eta_{R} \eta_{I},~Z \rightarrow \eta^{+} \eta^{-}$) and $W$ ($W \rightarrow \eta^{+} \eta_{R},~W \rightarrow \eta^{+} \eta_{I}$) to the odd sector particles,
\begin{equation}
\begin{split}
&  m_{\eta_{R}} + m_{\eta_{I}} > m_{Z},~~  2 m_{\eta^{+}} > m_{Z}, \\
& m_{\eta_{R}} + m_{\eta^{+}} > m_{W} ,~~  m_{\eta_{I}}+ m_{\eta^{+}} > m_{W}. 
\end{split}
\end{equation}

\item Production of $\eta^{+}$ at LEP II ($e^{+}~e^{-} \rightarrow \eta^{+} \eta^{-}$) also imposes strong constraints on the mass of $\eta^{+}$. Using the OPAL collaboration results \cite{OPAL:2003wxm,OPAL:2003nhx}, reference \cite{Pierce:2007ut} translated the constraints on charginos to the mass of $\eta^{+}$,   
\begin{equation}
m_{\eta^{+}} > 70~ \text{GeV}.
\end{equation}
\item Reference \cite{Lundstrom:2008ai} used the  DELPHI Collaboration study of neutralino pair production to constrain the masses of inert scalars. As a result, we excluded the region described by the intersection of the following conditions,
\begin{equation}
m_{\eta_{R}} < 80~ \text{GeV},~~~ m_{\eta_{I}} < 100 ~\text{GeV},~~~ m_{\eta_{I}}- m_{\eta_{R}} > 8~ \text{GeV}.
\end{equation}
This equation should be read as, if $m_{\eta_{R}}<80$ GeV and $m_{\eta_{I}}<100$ GeV then mass splitting ($m_{\eta_{I}}-m_{\eta_{R}}$ ) should not be greater than 8 GeV. Hence $m_{\eta_{R}}<80$ GeV are allowed if we follow, $m_{\eta_{I}}>100$ GeV and mass spllitting can be anything or $m_{\eta_{I}}<100$ GeV and mass splitting is less than 8 GeV.
\end{itemize} 
\item We also impose constraints coming from Higgs width measurements by LHC and CMS. The presence of additional states can be deduced indirectly from an increase in the width of the Standard Model Higgs. They will contribute to the invisible Higgs decays \cite{Barbieri:2006dq,Mandal:2021yph}. In particular, if $\eta_{R}$ and $\eta_{I}$ are light enough, there are two more decay channels for the SM-like Higgs boson,
\begin{equation}
\begin{split}
\Gamma_{h\rightarrow \eta_{R} \eta_{R}} =& \frac{v_{\Phi}^2 \lambda_{345}^{2}}{32 \pi m_{h}} \sqrt{1-\frac{4 m_{\eta_{R}}^{2}}{m_{h}^{2}}},\\
\Gamma_{h\rightarrow \eta_{I} \eta_{I}} =& \frac{(m_{\eta_{I}}^{2} - m_{\eta_{R}}^{2} + \frac{\lambda_{345}}{2}v_{\Phi}^{2}  )^{2}}{8 \pi v_{\Phi}^{2} m_{h}} \sqrt{1-\frac{4 m_{\eta_{I}}^{2}}{m_{h}^{2}}}.
\end{split}
\end{equation}
Note that due to the LEP constraints mentioned above ($m_{\eta^{+}}> 70~\text{GeV}$), there is no phase space for the two body decays $h \rightarrow \eta^{+} \eta^{-}$. These invisible decay modes of Higgs are constrained by the LHC and CMS experiments \cite{CMS:2018yfx}.
\begin{equation}
\text{BR}_{(h~\rightarrow~ \text{Inv} )} \leq 0.19.
\end{equation}
The value used for Higgs decay width in the Standard Model is $\Gamma_{h\rightarrow \text{SM}} \approx 4.07$~MeV.
\item For the case of scalar dark matter $\eta_R$ we fix $\lambda_4 +\lambda_5 <0$ and $\lambda_5 <0$ such that $m_{\eta^R} < m_{\eta^I},\,m_{\eta^{\pm}}$. We also ensure that $m_{\eta^R} < m_{N_{1,2}}$.
\item On the other hand for fermionic dark matter case, we make sure that $m_{N_1} < m_{\eta^R}, \,m_{\eta^I},\,m_{\eta^{\pm}}$.
\end{itemize}

\begin{figure}
	\includegraphics[scale=0.65]{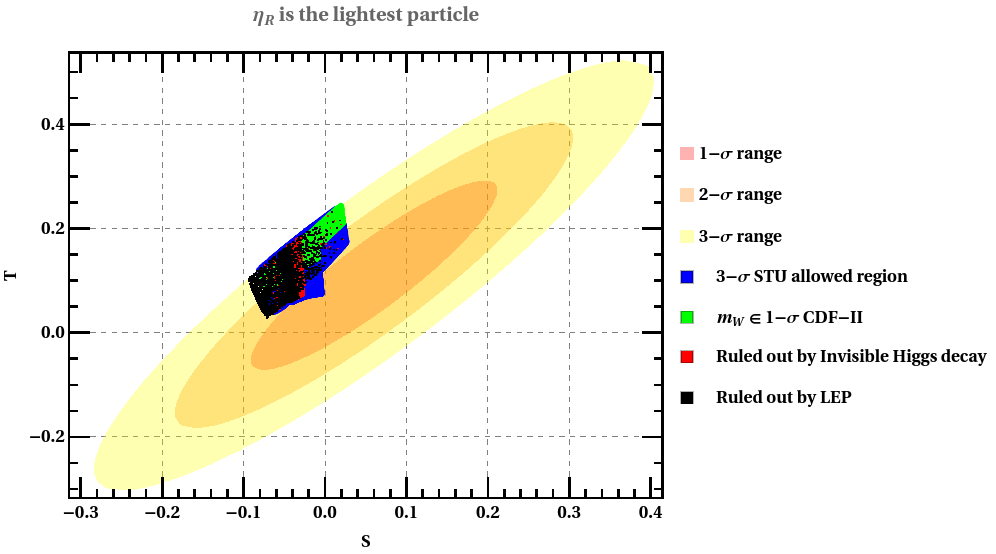}
	\caption{$S$ vs $T$ ellipse and $U \neq 0$ for the Scotogenic model. $\eta_{R}$ is taken as the lightest particle in the dark sector. As shown, the red region is ruled out by the Invisible Higgs decay constraint. The black points are ruled out by the LEP, the green region satisfies mass of $W$ boson $m_W$ at $1 $-$\sigma$ CDF-II, while the dark blue region shows the whole parameter space that is accesible by the Scotogenic model.}
	\label{fig:STU2}	
\end{figure}

\begin{figure}
	\includegraphics[scale=0.65]{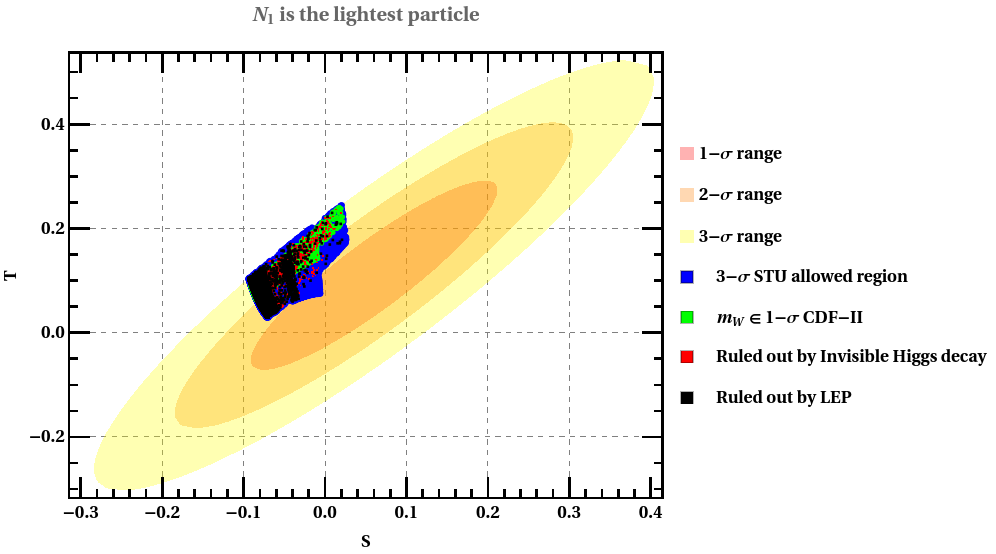}
	\caption{$S$ vs $T$ ellipse and $U \neq 0$ for the Scotogenic model. $N_{1}$ is taken as the lightest particle in the dark sector. The color code are same as Fig. \eqref{fig:STU2}.}
	\label{fig:STU3}	
\end{figure}

After the imposition of all the aforementioned constraints, the allowable parameter space in the $S$-$T$ plane under the conditions of $\eta_{R}$ being the lightest particle is illustrated in Fig.~\ref{fig:STU2} and $N_{1}$ being the lightest particle in Fig.~\ref{fig:STU3}. The three ellipses correspond to the 1-$\sigma$, 2-$\sigma$, 3-$\sigma$ regions of $S$ vs $T$ allowed by the new CDF-II measurement. The green zone corresponds to the region where the CDF-II mass of the $W$ boson $m_W$ is satisfied at $1 $-$\sigma$ along with all the aforementioned constraints, dark blue region displays the whole parameter space that the Scotogenic model can access. The red points is ruled out by Invisible Higgs decay constraints and the black are ruled out by LEP.

\section{The Scotogenic Dark matter}
\label{sec:drak-matter}
In this section, we study both the cases of scalar ($\eta_R$) and fermion ($N_1$) dark matter. We perform a detailed numerical scan for the model parameters with various experimental and theoretical constraints mentioned in previous sections. To do this analysis we  have implemented the model in SARAH~\cite{Staub:2015kfa} to calculate all the vertices, mass matrices, tadpole equations, whereas the thermal component of the dark matter relic, as well as dark matter direct detection cross sections are determined by micrOMEGAS-5.0.8~\cite{Belanger:2014vza}. 
\par In the following section we further use this allowed parameter space to check whether they can give correct relic abundance for scalar or fermionic dark matter as well as pass the dark matter direct detection constraints from experiments.

\subsection{Scalar dark matter $\eta_R$}
\begin{figure}[!htbp]
	\includegraphics[scale=0.27]{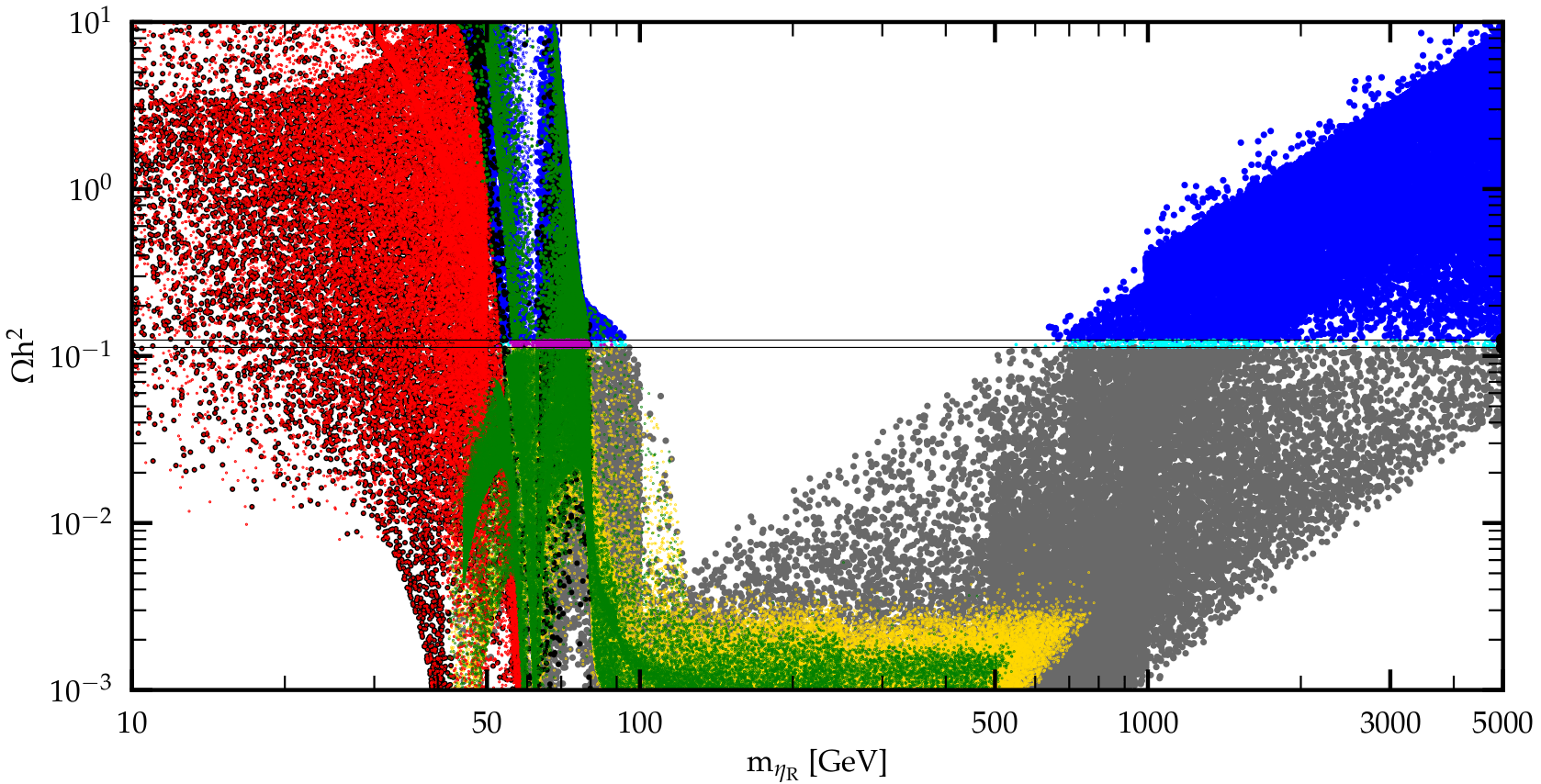}
	\caption{Relic density for the dark matter candidate $\eta_R$ for varying values of  $m_{\eta_R}$. Every point in the graph satisfies the lepton flavour violation bounds, vacuum stability bounds, and neutrino Yukawa bounds. The under and over-abundant relic density points, which do not satisfy CDF II $W$ mass, are shown by the gray and blue points, respectively. The cyan points satisfy all the constraints but not the revised $S$, $T$, and $U$ at 3$\sigma$ after the CDF II result. The red points are ruled out by invisible Higgs decay, and the black are ruled out by the LEP constraints. The yellow points satisfy all the constraints including $S$, $T$, and $U$ at 3$\sigma$, but not CDF II, $W$ boson mass measurement, and they are under abundant relic density points. Finally, the green points satisfy all the above constraints, including $S$, $T$, and $U$ at 3$\sigma$ and CDF II, $W$ boson mass measurement at 1$\sigma$. Magenta points are a subset of green points that also satisfy the experimental value of relic density at $3\sigma$.}
	\label{fig:relicscalar}
\end{figure}

We begin with the case of $\eta_R$ being the dark matter candidate. We show in Fig.~\ref{fig:relicscalar}, the behaviour of relic density of the scalar dark matter candidate $\eta_R$ as a function of the mass. The narrow band is the $3$-$\sigma$ allowed range for relic density : $0.1126 \leq \Omega_{\eta_{R}} h^2 \leq 0.1246$~\cite{Planck:2018vyg}. Our numerical scan was performed varying the input parameters as given in Table.~\ref{tab:scalar-dark-matter}, assuming logarithmic steps. The masses of the charged component of $\eta$ and the fermions $ N_1 $ and $ N_2 $ are taken in the range
\begin{equation}
	1 < \frac{m_{\eta^+}}{m_{\eta_0}}, \frac{M_{N_1}}{m_{\eta_0}}, \frac{M_{N_2}}{m_{\eta_0}} < 1000 .
\end{equation}
\par In scalar DM analysis we maintained a hierarchy $m_{\eta_{0}} < M_{N_{1}} < M_{N_{2}}$. The dips in the relic density plot in Fig~\ref{fig:relicscalar} can be understood from annihilation and co-annihilations diagrams shown in Fig. \ref{fig:ScalarAnnihilation}. The first dip at  $m_{\eta_R} \sim m_W /2 $ corresponds to the annihilation via s-channel $W$-exchange while the closely following second dip at $m_{\eta_R} \sim m_Z /2 $ corresponds to the annihilation and co-annihilation via s-channel $Z$-exchange. Similarly, the third dip at $m_{\eta_R} \sim m_h/2 \sim \;62~ \text{GeV}$ comes from the efficient annihilations through s-channel Higgs exchange.  The fourth dip at around $80$ GeV corresponds to the annihilation of $\eta_R$ to $W^+W^-$ and $Z Z$ through quartic $\eta_{R} \eta_{R}VV~(V=Z,W) $ interactions.

 
\par In Fig.~\ref{fig:relicscalar}, the under and over-abundant relic density points, which do not satisfy CDF II $W$ mass, are shown by the gray and blue points, respectively. The cyan points satisfy all the constraints but not the revised S,T, and U at 3$\sigma$ after the CDF II result. The red points are ruled out by invisible Higgs decay, and the black are ruled out by the LEP constraints. The yellow points satisfy all the constraints, including S, T, and U at 3$\sigma$, but not CDF II, W boson mass measurement, and they are under abundant relic density points. Finally, the green points satisfy all the above constraints including S, T, and U at 3$\sigma$ and CDF II, W boson mass measurement at 1$\sigma$. Magenta points are a subset of green points that also satisfy the experimental value of relic density at $3\sigma$.
\par There are two distint region of the dark matter $\eta_{R}$ mass that satisfies relic constraints. The low mass region, $54 ~\text{GeV}\lesssim m_{\eta_{R}} \lesssim 76~\text{GeV} $, and the high mass region, $m_{\eta_{R}} \gtrsim 500~ \text{GeV}$. As previously mentioned, large quartic couplings are necessary in the region with high dark matter mass in order to meet $S$, $T$, and $U$. Hence $m_{\eta_{R}} \gtrsim 500~ \text{GeV} $ is ruled out by the revised $S$, $T$, and $U$ values after CDF II results.
\par In the low mass region, $54 ~\text{GeV}\lesssim m_{\eta_{R}} \lesssim 76~\text{GeV} $, the dominant DM annihilation channels are Higgs resonance ($54 ~\text{GeV}\lesssim m_{\eta_{R}} \lesssim 62.5~\text{GeV} $), $\eta_{R}-\eta_{I}$ co-annihilation ($54 ~\text{GeV}\lesssim m_{\eta_{R}} \lesssim 71~\text{GeV} $) and annihilation of $\eta_R$ to $W^+W^-$ and $Z Z$ through quartic $\eta_{R} \eta_{R}VV~(V=Z,W) $ interactions ($71~\text{GeV}\lesssim m_{\eta_{R}} \lesssim 76~\text{GeV} $). Note that in this mass region of $\eta_{R}$, LEP restricts the mass splitting between neutral scalars to be less than 8 GeV for $m_{\eta_{I}} < 100~\text{GeV}$. Hence in the co-annihilation dominant region, to satisfy the relic density, the mass splitting between neutral scalars is in the range of 7 GeV to 8 GeV, and the mass of $\eta^{+}$ is in the range of 120 GeV to 200 GeV.

\begin{figure}[!htbp]
	\includegraphics[scale=0.35]{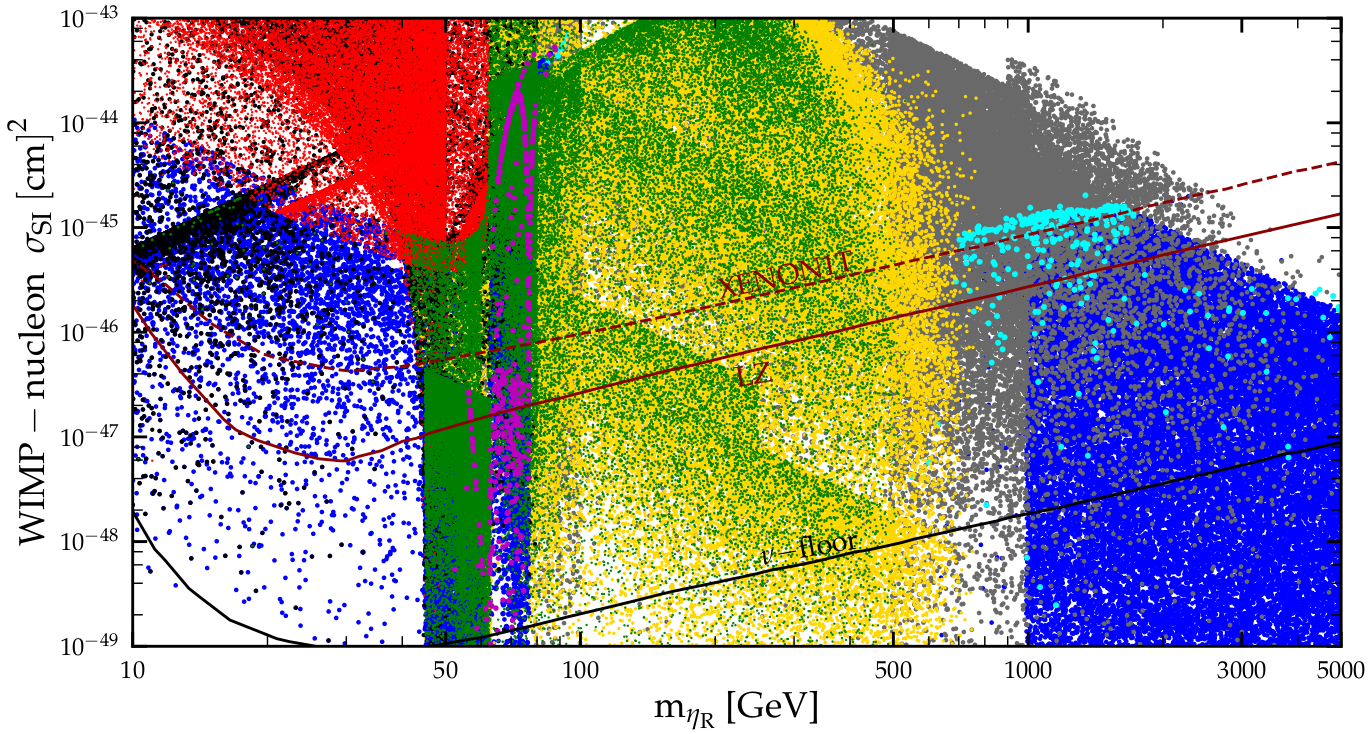}
	\caption{Spin-independent WIMP-nucleon cross section for the dark matter candidate $\eta_R$ vs $m_{\eta_R}$. The color scheme is the same as in Fig.~\ref{fig:relicscalar}. The dark solid red line denotes the latest upper bound from the LZ \cite{LUX-ZEPLIN:2022qhg} collaboration. The dark dashed red line denotes the  upper bound from the XENON1T collaboration~\cite{XENON:2018voc}, and the solid black line corresponds to the “neutrino floor” lower limit~\cite{Billard:2013qya,Billard:2021uyg}.}
	\label{fig:DDscalar}
\end{figure}

The scalar dark matter $\eta_R$ can interact with nucleons through tree level Higgs $h$ and $Z$ boson exchanges as show in Fig.~\ref{fig:direct} in Appendix B.  Thus, for the scalar dark matter case, the dark matter direct detection experiments can put stringent constraints on the allowed parameter space. There are constraints from various direct detection experiments such as XENON1T~\cite{XENON:2018voc}, LUX-ZEPLIN (LZ)\cite{LUX-ZEPLIN:2022qhg,A:2022acy}, DEAP~\cite{Lehnert:2018vzs}, LUX~\cite{LUX-ZEPLIN:2018poe}, WARP~\cite{Zani:2014lea},  and CDMS~\cite{CDMS-II:2009ktb}. Recent results from the LUX-ZEPLIN (LZ) collaboration's impose the most stringent constraints. However, we plotted both XENON1T and LZ constraints in Fig.~\ref{fig:DDscalar} for comparison. Indirect dark matter detection experiments, such as IceCube \cite{IceCube:2016dgk}, Super-K \cite{Frankiewicz:2015zma}, and Antares \cite{ANTARES:2016xuh}, provide constraints on the mass and cross section of dark matter. However, these constraints are relatively weaker compared to the results obtained from the LZ and XENON1T experiments \cite{deBoer:2021pon}.
\par The Higgs mediated DM-nucleon interaction is governed by $\lambda_{345}$ coupling. To satisfy the LZ bound on direct detection cross-section, we find that the range of $\lambda_{345}$ is $-0.006 \lesssim \lambda_{345} \lesssim 0.006$. In Fig.~\ref{fig:DDscalar}, we show the direct detection prospects of our dark matter candidate $\eta_R$, for the range of parameters covered by our scan given in Table \ref{tab:scalar-dark-matter}. The color code is same as of Fig.~\ref{fig:relicscalar}. The points above the dark solid red line are ruled out by LZ collaboration and the point above dark dashed red line is ruled out by the XENON1T collaboration. We also show the lower limit corresponding to the “neutrino floor” from coherent elastic neutrino scattering~\cite{Billard:2013qya,Billard:2021uyg}. We find that the low mass region,
\begin{equation}
54 ~\text{GeV}\lesssim m_{\eta_{R}} \lesssim 76~\text{GeV},
\end{equation}
is compatible with all the constraints and can satisfy the CDF II, $W$ boson measurement at 1$\sigma$.
\subsection{Fermionic dark matter $N_1$}
In this section, we study the fermionic dark matter candidate $N_1$~\cite{Abada:2018zra,Schmidt:2012yg}. The numerical scan was performed by varying the input parameters as given in Tab.~\ref{tab:fermion-dark-matter}, following the same constraints as mentioned above. The annihilation channels which determines the dark matter relic abundance are $N_1 N_1 \to \ell_i\ell_j$, $\nu_i\nu_j$ via the Yukawa couplings. In order to obtain the correct relic abundance, the magnitude of the Yukawa couplings should be relatively large. But such large Yukawa couplings are strongly constrained by bounds from LFV. However, these LFV constraints can be evaded when co-annihilation effects for dark matter relic abundance become important. This happens when $\eta^{\pm}$ or $\eta_{I}/\eta_{R}$ masses are close to $N_1$ mass. In Fig.~\ref{fig:FermionAnnihilation} and \ref{fig:FermionCoAnnihilation} we have listed all the relevant annihilation and co-annihilation channels for the fermionic dark matter $N_1$. Motivated from this discussion, we consider the following intervals in our numerical computations for fermionic dark matter:
\begin{equation}
1 < \frac{M_{N_2}}{M_{N_1}}, \frac{m_{\eta_0}}{M_{N_1}} < 10.
%
\label{eq:fermionicscan}
\end{equation}
\begin{table}[!htbp]
	\begin{center}
		\begin{tabular}{ | >{\centering\arraybackslash}m{1in} |>{\centering\arraybackslash}m{3in}| }
			\hline
			Parameter & Range \\
			\hline
			&\\[-12pt]
			$\lambda_2$  & $\bm{[} \, 10^{-6},  \sqrt{4 \pi} \, \bm{]} \; $ \\
			$\lambda_3$  & $\bm{[} \,  10^{-6},  \sqrt{4 \pi}  \, \bm{]} \; $ \\ 
			$\lambda_4$  & $\bm{[} \,-\sqrt{4 \pi},  -10^{-6} \, \bm{]} \; $ \\
			$\lambda_5$  & $\bm{[} \, -\sqrt{4 \pi},  -10^{-6}\, \bm{]} \; $ \\
			$M_{N_1}$  & $\bm{[} \, 10, 5000\;   \bm{]} \text{GeV} $ \\
			$M_{N_2}$  & $\bm{[} \, M_{N_1} + \delta, 10 M_{N_1} \, \bm{]} \text{GeV} $ \\
			$m_{\eta^0}$  & $\bm{[} \, M_{N_1} + \delta, 10 M_{N_1} \, \bm{]} \text{GeV} $ \\
			\hline
		\end{tabular}
	\end{center}
	\caption{Value range for the numerical parameter scan for fermion dark matter. Here, $\delta$ represents a small positive number. Within this range, we have ensured that $M_{N_{1}}$ is less than $M_{N_{2}}$ and $m_{\eta_{0}}$.}
	\label{tab:fermion-dark-matter}
\end{table}

\begin{figure}[!htbp]
	\includegraphics[scale=0.27]{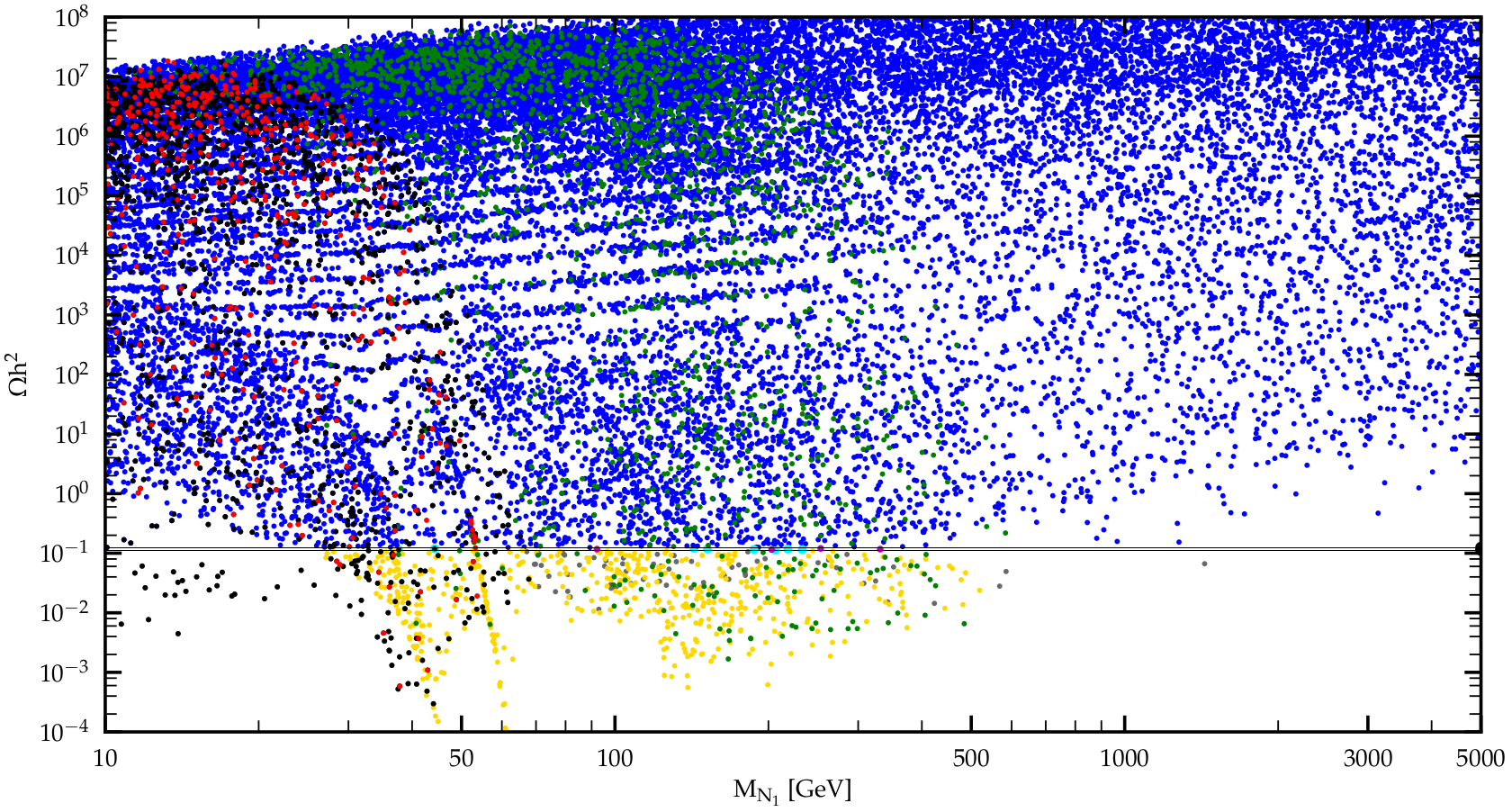}
	\caption{Relic density for the fermionic dark matter candidate $N_1$ as a funcion of mass $M_{N_1}$ varying the parameters as given in Eq.~\ref{eq:fermionicscan}. The color codes are same as in Fig.~\ref{fig:relicscalar}.}
	\label{fig:relicfermionlarge}
\end{figure}

\begin{figure}[!htbp]
	\includegraphics[scale=0.27]{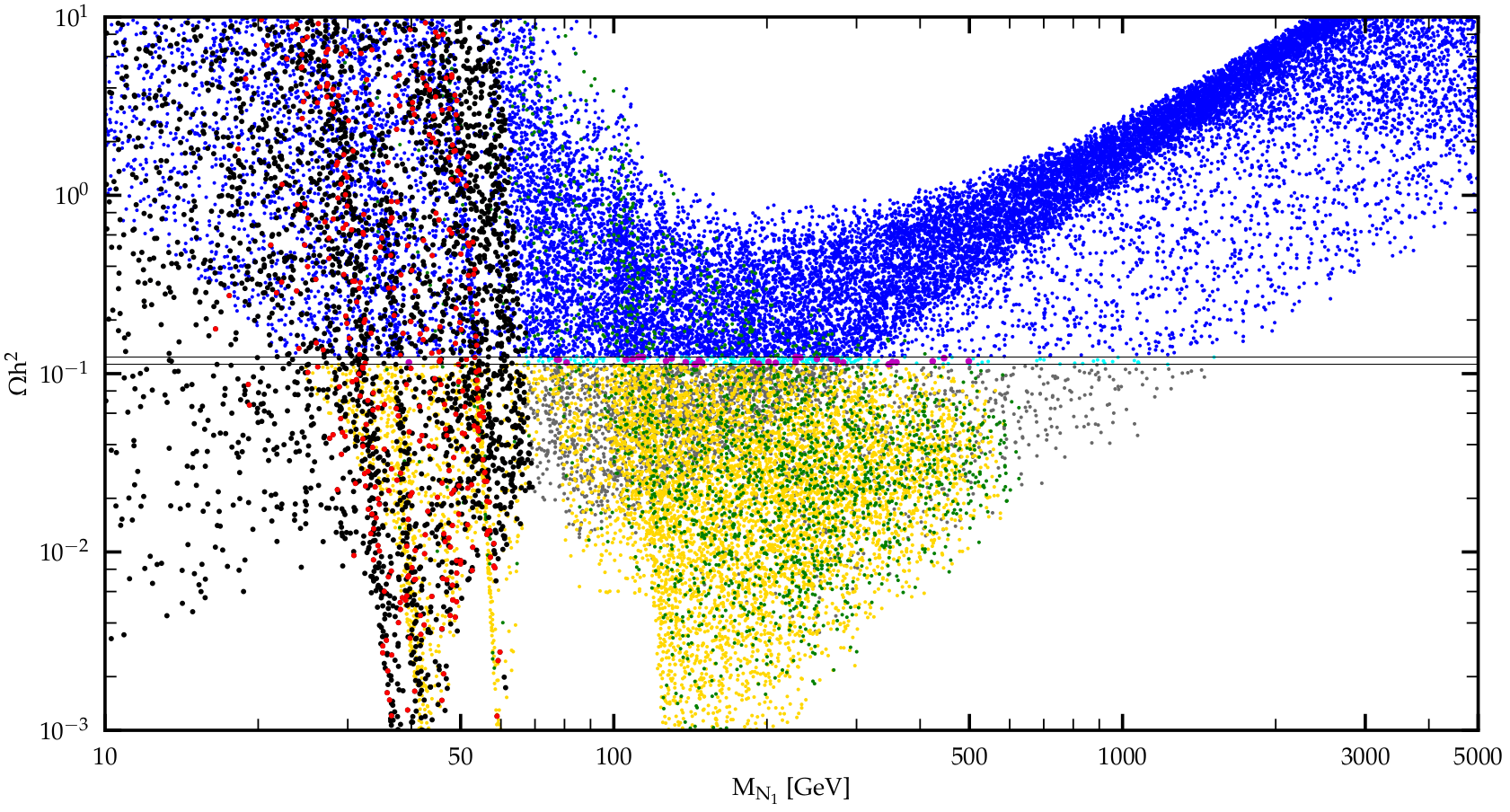}
	\caption{Relic density for fermionic dark matter candidate $N_1$ as function of mass $M_{N_1}$ varying the parameters as given in Eq.~\ref{eq:fermionicscan2}. The color codes are same as in Fig.~\ref{fig:relicscalar}.}
	\label{fig:relicfermion}
\end{figure}

In Fig.~\ref{fig:relicfermionlarge}, we show the behaviour of relic density for fermionic dark matter $ N_1 $ using the mass variation as given in Eq.~\ref{eq:fermionicscan}. We see that we are getting very few points with correct relic density as for most of the points the mass ratios $\frac{m_{\eta^0}}{M_{N_1}}$ are not close to $\mathcal{O}(1)$. As a result, co-annihilation is not efficient, and the low value of Yukawa required to meet the LFV constraints suppresses the annihilation channels as well. For this reason, in order to obtain correct relic abundance we have to consider the parameter space where co-annihilation diagrams shown in Fig.~\ref{fig:FermionCoAnnihilation} become important. This happens when we consider small mass differences between $ N_1 $ and $ \eta_{R} $, $ \eta_{I} $ and $ \eta_{\pm} $. Hence we vary the parameters in the range
\begin{equation}
M_{N_1} < M_{N_2}, m_{\eta^0} < M_{N_1} + 10\,\text{GeV}.
\label{eq:fermionicscan2}
\end{equation}
In Fig.~\ref{fig:relicfermion} we plot the relic density with respect to the  fermionic dark matter mass $M_{N_1}$ after imposing the condition given in Eq.~\eqref{eq:fermionicscan2}. 
From Fig.~\ref{fig:relicfermion} we can see that in this case, co-annihilation processes given in Fig.~\ref{fig:FermionCoAnnihilation} dominate over annihilation channels. The efficient co-annihilation of the fermionic dark matter in turn leads to larger cross-sections, resulting in many points in Fig.~\ref{fig:relicfermion} which satisfy relic density constraints.
Finally, before ending this section we note that in this case there is no significant limitation from XENON1T and LZ direct detection constraints because the fermionic dark matter candidate $N_1$ interacts exclusively with leptons through the Yukawa coupling $Y_N$ and does not interact with quarks or gluons at tree level. We find the fermionic dark matter candidate in the range $M_{N_1} < 500$ GeV, can
simultaneously explain all the aforementioned constraints we discussed earlier.



\section{Dark matter analysis in view of ATLAS 2023 Result}
\label{ATLAS}

ATLAS released their findings in 2017 regarding the measurement of the W boson mass $80.370 \pm 0.019$ GeV \cite{ATLAS:2017rzl}, using proton-proton collision data. The data was collected in 2011 at the LHC, with a center-of-mass energy of 7 TeV. The dataset corresponds to an integrated luminosity of $4.6~\text{fb}^{-1}$. They further reanalysed their 2011 sample of W boson and published their result on 23 March 2023. They improved the precision of their previous measurement and the new W boson mass by ATLAS is $80.360 \pm 0.016$ GeV ~\cite{ATLAS:20233}.The present measurement from ATLAS are consistent with their 2017 findings and aligns with the value predicted by the SM. 
\par To see the effect of ATLAS 2023 resluts we re-analysed the data we previously generated. Ref. \cite{Lu:2022bgw} showcased global electroweak fits using two distinct values of the $W$ boson mass as input parameters. The study used the $W$ mass values from both CDF II and PDG 2021, providing a comparative analysis. They found following values of $S, T$ and $U$ parameters: \\\\   
\underline{PDG 2021 :} Oblique parameters at $1 \sigma$
\begin{equation}
\begin{split}
S=0.06 \pm 0.10,~T=0.11 \pm 0.12,~U=-0.02 \pm 0.09.\\
\text{with the correlation}~~\rho_{ST}=0.9,~\rho_{SU}=-0.57,~\text{and}~\rho_{TU}=-0.82.
\end{split}
\end{equation} 
\underline{CDF 2022 :} Oblique parameters at $1 \sigma$
\begin{equation}
\begin{split}
S=0.06 \pm 0.10,~T=0.11 \pm 0.12,~U=0.14 \pm 0.09.\\
\text{with the correlation}~~\rho_{ST}=0.9,~\rho_{SU}=-0.59,~\text{and}~\rho_{TU}=-0.85.
\end{split}
\end{equation}    
\par The primary distinctions beween PDG 2021 and CDF II lie in the significantly larger central value of the $U$ parameter predicted by CDF 2022 compared to 2021, along with a slight strengthening of the correlations between $U$ and $S$ or $T$. Our approach involves revisiting the data we previously generated and incorporating the oblique parameters obtained through an electroweak fit using the $W$ boson mass from PDG (2021) into our analysis. In addition we will make sure that the new result of ATLAS on $W$ boson mass is also satisfied. 
\begin{figure}[ht]
   \centering
   \captionsetup{justification=raggedright}
  \includegraphics[width=0.95\textwidth]{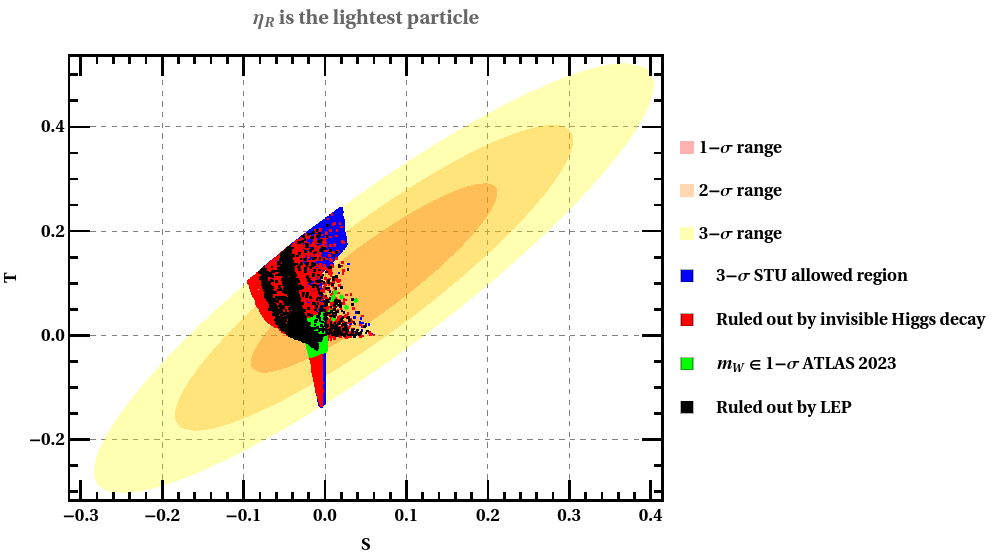}
  \caption{ $S$ vs $T$ ellipse with $U \neq 0$ for the Scotogenic model. $\eta_{R}$ is taken as the lightest particle in the dark sector. As shown, the red region is ruled out by the Invisible Higgs decay constraint. The black points are ruled out by the LEP, the green region satisfies mass of $W$ boson $m_{W}$ at $1\sigma$ ATLAS 2023, while the dark blue region shows the whole parameter space that is accesible by the Scotogenic model.}
  \label{STUnew}
\end{figure}
Fig. \ref{STUnew} illustrates the permissible parameter space in the $S-T$ plane, considering the constraints mentioned earlier, with $\eta_{R}$ being the lightest particle. The three ellipses represent the $1\sigma$, $2\sigma$, and $3\sigma$ regions of $S$ versus $T$ allowed by the electro weak fit using PDG 2021, W boson mass. The green zone corresponds to the region where the ATLAS 2023 W boson mass satisfies the $1\sigma$ criterion, along with all the aforementioned constraints. The dark blue region represents the entire parameter space accessible to the Scotogenic model. The red points are excluded by the constraints of Invisible Higgs decay, while the black points are excluded by LEP. The primary difference between  Fig. \ref{fig:STU2} and Fig. \ref{STUnew} is the significant large value of $T$ parameter in Fig. \ref{fig:STU2}. As the $T$ parameter dominates the $W$ mass corrections, a large $T$ value is required to satisfy the CDF-II results. Hence CDF-II, $W$ boson mass at $1\sigma$ can only be satisfied in the $3\sigma$ zone of the $(S, T)$ plane. In the ATLAS-2023 case, $W$ boson mass at $1\sigma$ can even be satisfied within the $1\sigma$ zone of the $(S, T)$ plane. Small negative $T$ values are also allowed in this case.   

\par
\begin{figure}[ht]
   \centering
   \captionsetup{justification=raggedright}
  \includegraphics[width=0.99\textwidth]{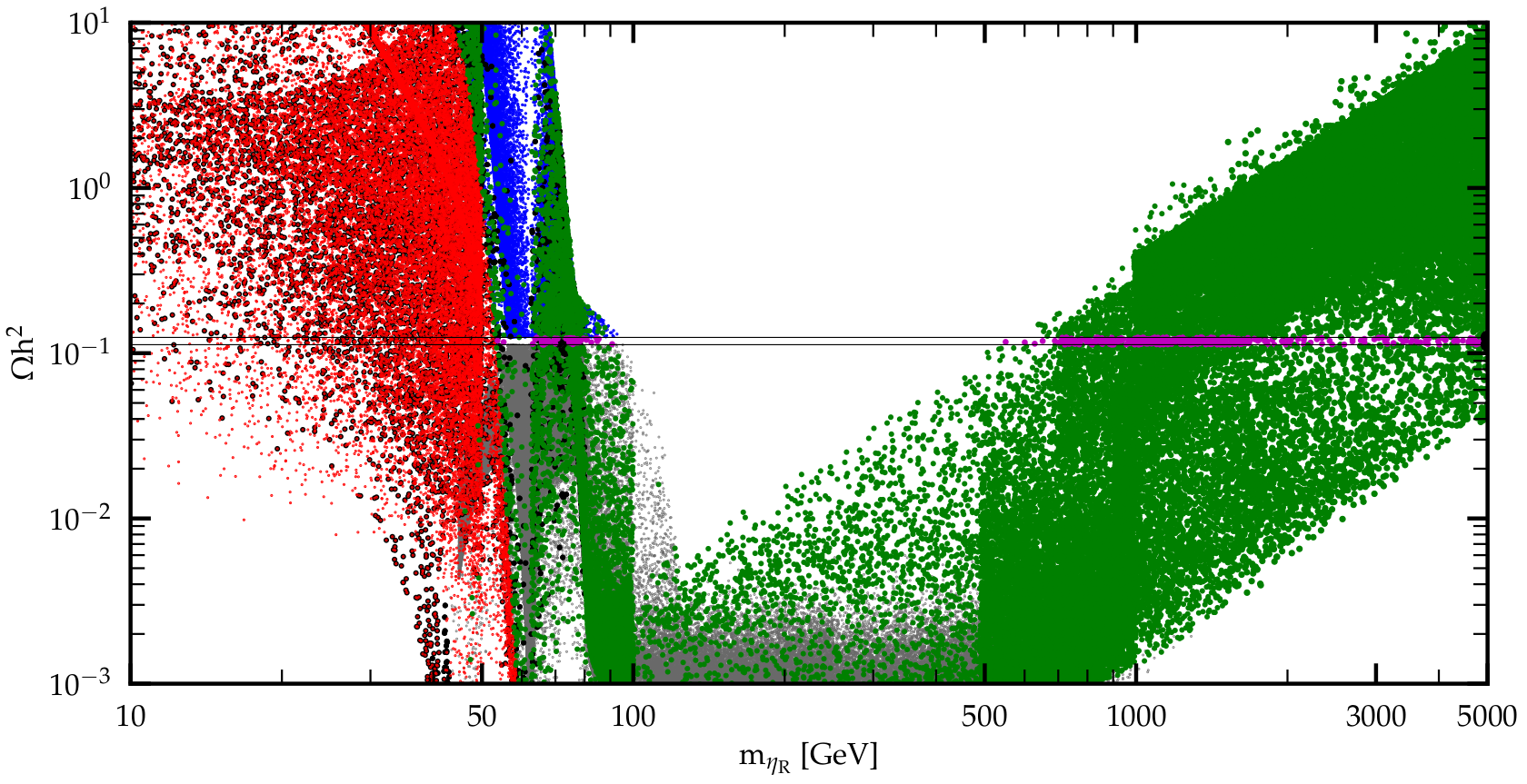}
  \caption{ Relic density for the dark matter candidate $\eta_{R}$ for varying values of $m_{\eta_{R}}$ . Every point in the graph satisfies the lepton flavour violation bounds, vacuum stability bounds, neutrino Yukawa bounds, and $S,T,U$ values allowed by the electro weak fit using PDG 2021, $W$ boson mass. The under and over-abundant relic density points, which do not satisfy ATLAS 2023 result, are shown by the gray and blue points, respectively. The red points are ruled out by invisible Higgs decay, and the black are ruled out by the LEP constraints. Finally, the green points satisfy all the above constraints and ATLAS 2023, $W$ boson mass measurement at $1\sigma$. Magenta points are a subset of green points that also satisfy the experimental value of relic density at $3\sigma$.}
  \label{RelicNew}
\end{figure}

 To see the effect on the dark matter constraints, we redraw the Fig. \ref{fig:relicscalar}. Fig. \ref{RelicNew} depicts the relic density for the dark matter candidate, $\eta_{R}$. Each point on the graph satisfies the bounds coming from lepton flavor violation, vacuum stability, neutrino Yukawa, and the allowed $S,T,U$ values determined by the electroweak fit using PDG 2021 $W$ boson mass. Points representing under-abundant relic density, which do not align with the ATLAS 2023 result, are shown in gray, while over-abundant relic density points are shown in blue. Red points are excluded due to invisible Higgs decay, and black points are excluded by LEP constraints. On the other hand, green points satisfy all the aforementioned constraints as well as the ATLAS 2023 $W$ boson mass measurement within $1\sigma$ uncertainty. Furthermore, magenta points are a subset of the green points, satisfying both the experimental relic density value within $3\sigma$ uncertainty.

\par The difference between Fig.~\ref{RelicNew} and the Fig.~\ref{fig:relicscalar} lies in the treatment of the high mass region ($m_{\eta_{R}} \gtrsim 500$ GeV). Fig.~\ref{RelicNew} allows for the inclusion of this region when considering ATLAS measurements, whereas the preliminary Fig.~\ref{fig:relicscalar} excludes it based on CDF II measurements. Notably, all the cyan points in the Fig.~\ref{fig:relicscalar} remain in agreement with the ATLAS 2023 measurements. This result is also in agreement  with all previous analyses conducted on the inert doublet model and Scotogenic model. Fig.~\ref{fig:relicfermionlarge} and Fig.~\ref{fig:relicfermion} for fermionic dark matter exhibit comparable modifications, resulting in the inclusion of cyan points that were previously ruled out by the CDF II measurement. These points will now be allowed based on the ATLAS 2023 measurement.
\section{Conclusions}
\label{sec:conclusion}
The canonical Scotogenic model is one of the simplest models which can explain both small neutrino masses and dark matter together. In the Scotogenic model, neutrino mass generation is intimately related with the existence of dark matter, with dark sector particles running in the loop diagram for neutrino mass generation. Despite its simplicity and elegance, the model is rich and flexible enough to explain the recent CDF-II measurement of $W$ boson mass which shows 7-$\sigma$ deviation from the Standard Model expectation. In this work we have shown that the new physics corrections to $W$ boson mass in the canonical scotogenic model come primarily from the loop corrections due to the presence of a dark $SU(2)_L$ doublet scalar $\eta$. For both cases of scalar as well as fermionic dark matter, we have analyzed the self-consistency of the model by simultaneously demanding that:
\begin{itemize}
 \item The model satisfies the $S$, $T$, $U$ parameters consistent with CDF-II measurement.
 \item It has couplings consistent with stability and perturbativity constraints.
 \item All neutrino physics constraints are satisfied i.e. those coming from neutrino oscillation experiments as well as the limits from cosmology, KATRIN and neutrinoless double beta decay experiments.
 \item The model satisfies the constraints from lepton flavour violation.
 \item  Constraints coming from LEP and LHC are satisfied. 
 \item The constraints coming from dark matter direct detection and relic abundance are also satisfied.
\end{itemize}

We demonstrate that the new CDF-II measurement rules out the feasible parameter space of a doublet scalar in high mass regions ($m_{\eta_{R}} \gtrsim 500~\text{GeV}$) while still allowing the low mass regions of ($54~\text{GeV} \lesssim m_{\eta_{R}} \lesssim 76~\text{GeV}$).  We also showed that the fermion dark matter ($M_{N_{1}} \lesssim 500~\text{GeV} $) has enough parameter space to simultaneously satisfy all the aforementioned constraints. 
Additionally, we demonstrated that the parameter space previously excluded by CDF II is now deemed permissible based on the latest results obtained by ATLAS.


\begin{acknowledgments}
The work of RS is supported by the Government of India, SERB Startup Grant SRG/2020/002303. The work of S.M. is supported by KIAS Individual Grants (PG086001) at Korea Institute for Advanced Study.
\end{acknowledgments}

\appendix

\section{Definition of $S$, $T$ and $U$ parameters}\label{APA}
Following closely the notation by Peskin and Takeuchi in \cite{Peskin:1991sw}, the $S, T$ and $U$ parameters can be defined as
\begin{subequations}
\begin{align}
\alpha \, S &\equiv 4 e^{2} \: \frac{d}{dp^{2}} \left [\Pi_{33}(p^{2}) - \Pi_{3Q}(p^{2})
\right ] \biggl |_{p^{2}=0} \;, \\
\alpha \, T &\equiv \frac{e^{2}}{s_{W}^{2} c_{W}^{2} m_{Z}^{2}}\: 
\left [\Pi_{11}(0) - \Pi_{33}(0) \right ]\;, \\
\alpha \, U &\equiv 4 e^{2} \: \frac{d}{dp^{2}} \left [\Pi_{11}(p^{2}) - \Pi_{33}(p^{2})
\right ]\biggl |_{p^{2}=0} \;,  
\end{align}
\label{eq:STU}
\end{subequations}
where $\alpha \equiv \alpha_{em} = e^{2}/4\pi$. $\Pi_{IJ}\equiv \Pi_{IJ}(p^{2})$ are the various vaccum polarisation diagrams
where $I$ and $J$ may be photon ($\gamma)$, $W$ or $Z$ boson, 
\begin{subequations}
\begin{align}
\Pi_{\gamma\gamma} &= e^{2}\: \Pi_{QQ} \;, \\
\Pi_{Z\gamma} &= \frac{e^{2}}{c_{W}s_{W}} \: \left ( \Pi_{3Q} - s^{2} \Pi_{QQ} \right )\;,\\
\Pi_{ZZ} &= \frac{e^{2}}{c_{W}^{2} s_{W}^{2}} \: 
\left (\Pi_{33} - 2 s^{2} \Pi_{3Q} + s^{4} \Pi_{QQ}\right )\;,\\
\Pi_{WW} &= \frac{e^{2}}{s_{W}^{2}} \: \Pi_{11} \;,
\end{align}
\end{subequations}
where $\theta_{W}$ is the weak mixing angle and $s_{W}=\sin\theta_{W}, c_{W}=\cos\theta_{W}$.\\
For scotogenic model, these oblique parameters are given as,
\begin{equation}\label{Tpara}
\begin{split}
T=& \frac{1}{16\pi^2 \alpha_{em}v_{\Phi}^2} \left[ \sum_{i=1}^{3} (1-T_{1i}^2)F(m_{\eta^{+}}^{2},m_{i}^{2}) - T_{11}^{2}F(m_{2}^{2},m_{3}^{2})-T_{12}^{2}F(m_{3}^{2},m_{1}^{2})-T_{13}^{2}F(m_{1}^{2},m_{2}^2)  \right. \\
&~~~ \left.+3\sum_{i=1}^{3} T_{1i}^{2} \{F(m_{Z}^{2},m_{j}^{2})-F(m_{W}^{2},m_{j}^{2})\} -3 \{F(m_{Z}^{2},m_{h}^{2})-F(m_{W}^{2},m_{h}^{2})\} \right].
\end{split}
\end{equation}
Where,~ $T_{ij}=\delta_{ij}$,~ $m=\{m_{h}, m_{\eta_{R}}, m_{\eta_{I}} \}$,~ and
\begin{equation}
    F(x,y)= 
\begin{cases}
    \frac{x+y}{2}-\frac{xy}{x-y}\text{log}\left[\frac{x}{y}\right],& \text{if } x\neq y\\
    0,              & \text{otherwise}
\end{cases}.
\end{equation}
After simplyfing \eqref{Tpara} we have
\begin{equation} \label{Tscoto}
T=\frac{1}{16\pi^{2} \alpha_{em} v_{\Phi}^{2}} \left[ F(m_{\eta^{+}}^{2},m_{\eta_{R}}^{2})+F(m_{\eta^{+}}^{2},m_{\eta_{I}}^{2})-F(m_{\eta_{I}}^{2},m_{\eta_{R}}^{2})  \right].
\end{equation}
For expression of $S$ and $U$ we need to define two more functions:
\begin{equation}
\begin{split}
G(x,y)=& -\frac{16}{3}+5(x+y)-2(x-y)^{2} +3 \left[ \frac{x^{2}+y^{2}}{x-y} -x^{2}+y^{2} +\frac{(x-y)^{3}}{3} \right]\text{log} \frac{x}{y}\\
&~~ [1-2(x+y)+(x-y)^{2}]f(x+y-1,1-2(x+y)+(x-y)^{2}),\\
\end{split}
\end{equation}
\begin{equation}
\begin{split}
\tilde{G}(x,y)=& -\frac{79}{3}+9x-2x^{2} + \left[ -10 +18x-6x^{2}+x^{3}-9\frac{x+1}{x-1} \right]\text{log}\left(x\right) ~~~~~~~~~~~~~~~~\\
&~~~~~~~~~~~~~~~~~~~~~~~~~~~~~~~~~~~~~~~~~~~+~(12-4x+x^{2})f(x,x^2-4x).
\end{split}
\end{equation}
Where\\
\begin{equation}
    f(x,y)= 
\begin{cases}
    \sqrt{y}~\text{log}\left|\frac{x-\sqrt{y}}{x+\sqrt{y}}\right|,& \text{if }~ y\ge 0,\\
    0,              & \text{if}~ y=0,\\
   2\sqrt{-y}~\text{tan}^{-1}[\frac{\sqrt{-y}}{x}],              & \text{if}~ y\le 0\\ 
\end{cases}.
\end{equation}\\
With this definitions, $S$ and $U$ can be written as
\begin{equation}\label{Sscoto}
\begin{split}
S=& \frac{1}{24\pi}\biggl\{(s_{W}^{2}-c_{W}^{2})^{2}G(z_{\eta^{+}},z_{\eta^{+}})+T_{11}^{2}G(z_{\eta_{R}},z_{\eta_{I}}) + T_{12}^{2}G(z_{\eta_{I}},z_{h}) + T_{13}^{2}G(z_{h},z_{\eta_{R}}) \biggr. \\
&~~ \left.  + \sum_{i=1}^{3}\left[ T_{1i}^{2}\tilde{G}(z_{i}) + \text{log}\frac{m_{i}^{2}}{m_{\eta^{+}}^{2}}   \right] - \tilde{G}(z_{h}) -\text{log}\frac{m_{h}^{2}}{m_{\eta^{+}}^{2}}\right\},
\end{split}
\end{equation}

\begin{equation}\label{Uscoto}
\begin{split}
U=& \frac{1}{24\pi}\biggl\{ \sum_{i=1}^{3} (1-T_{1i}^{2})G(w_{\eta^{+}},w_{i}) -(s_{w}^{2}-c_{w}^{2})^{2}G(z_{\eta^{+}},z_{\eta^{+}})-T_{11}^{2}G(z_{\eta_{R}},z_{\eta_{I}}) - T_{12}^{2}G(z_{\eta_{I}},z_{h}) \biggr. \\
&~~ \left. - T_{13}^{2}G(z_{h},z_{\eta_{R}})+ \sum_{i=1}^{3}T_{1i}^{2}\left[ \tilde{G}(w_{i})- \tilde{G}(z_{i})   \right] - \tilde{G}(w_{h})+ \tilde{G}(z_{h}) \right\}.
\end{split}
\end{equation}
Where
\begin{equation}
z_{a}=\frac{m_{a}^{2}}{m_{Z}^{2}} ~\text{and}~w_{a}= \frac{m_{a}^{2}}{m_{W}^{2}},
\end{equation}
 $a=\{\eta^{+},h,\eta_{R},\eta_{I},h\}$ and $m=\{m_{h}, m_{\eta_{R}}, m_{\eta_{I}} \}$ .
 \par It is worth mentioning that under the assumption $m_{\eta_{R}} \approx m_{\eta_{I}}$, equations \eqref{Sscoto},\eqref{Tscoto} and \eqref{Uscoto} can be written as
\begin{equation}
S \approx \frac{1}{12\pi} \text{log}\left( \frac{m_{\eta_{R}}^{2}}{m_{\eta^{+}}^{2}} \right),~~ T \approx \frac{1}{8\pi^{2} \alpha_{em} v_{\Phi}^{2}}F(m_{\eta^{+}}^{2},m_{\eta_{R}}^{2}), \text{and}~~ U \approx \frac{1}{12\pi} G\left(\frac{m_{\eta^{+}}^{2}}{m_{W}^{2}}, \frac{m_{\eta_{R}}^{2}}{m_{W}^{2}}\right). 
\end{equation}

\section{Feynman diagrams for dark matter relic density and direct detection}
\label{app2-feynman}
\begin{figure}[!htbp]
	\centering
	\includegraphics[scale=1.5]{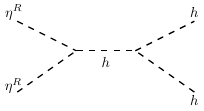}
	\includegraphics[scale=1.5]{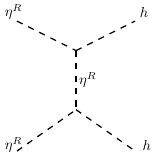}
	\includegraphics[scale=1.5]{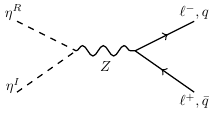}
	\includegraphics[scale=1.5]{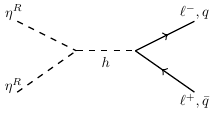}
	\includegraphics[scale=1.5]{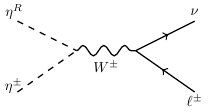}
	\includegraphics[scale=1.5]{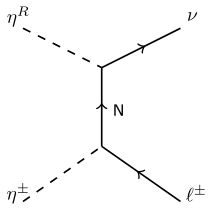}
	\includegraphics[scale=1.5]{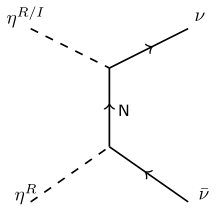}
	\includegraphics[scale=1.5]{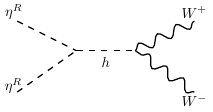}
	\includegraphics[scale=1.5]{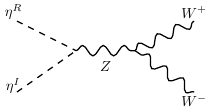}
	\includegraphics[scale=1.5]{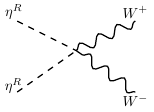}
	\includegraphics[scale=1.5]{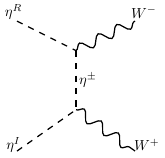}
	\includegraphics[scale=1.5]{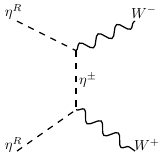}
	\includegraphics[scale=1.5]{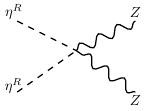}
	\caption{Annihilation and co-annihilation Feynman diagrams that contribute to the relic density of the scalar dark matter candidate $ \eta_R $.}
	\label{fig:ScalarAnnihilation}
\end{figure}

\begin{figure}[!htbp]
	\centering
	\includegraphics[scale=1.5]{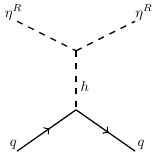}
	\includegraphics[scale=1.5]{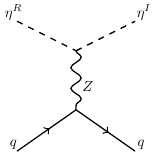}
	\caption{$ \eta_R $-nucleon scattering Feynman diagrams relevant for direct detection.}
	\label{fig:direct}
\end{figure}

\begin{figure}[!htbp]
	\centering
	\begin{subfigure}{\linewidth}
		\includegraphics[scale=0.28]{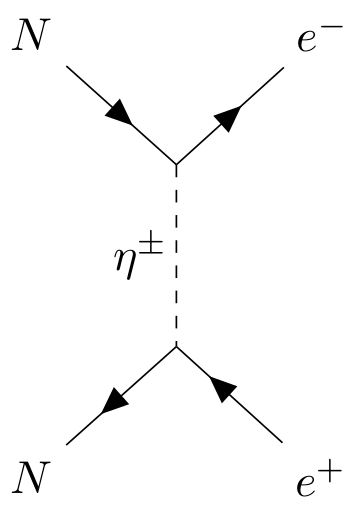}
		\includegraphics[scale=0.28]{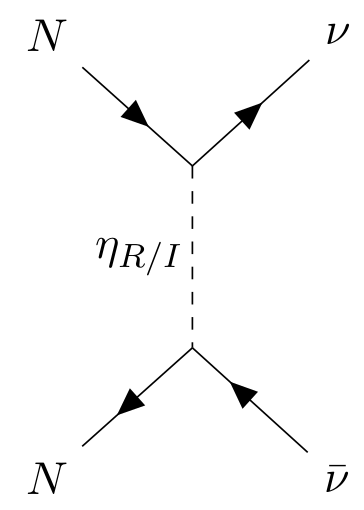}
		\caption{\centering Annihilation Channels}
		\label{fig:FermionAnnihilation}
	\end{subfigure}
	\begin{subfigure}{\linewidth}
		\includegraphics{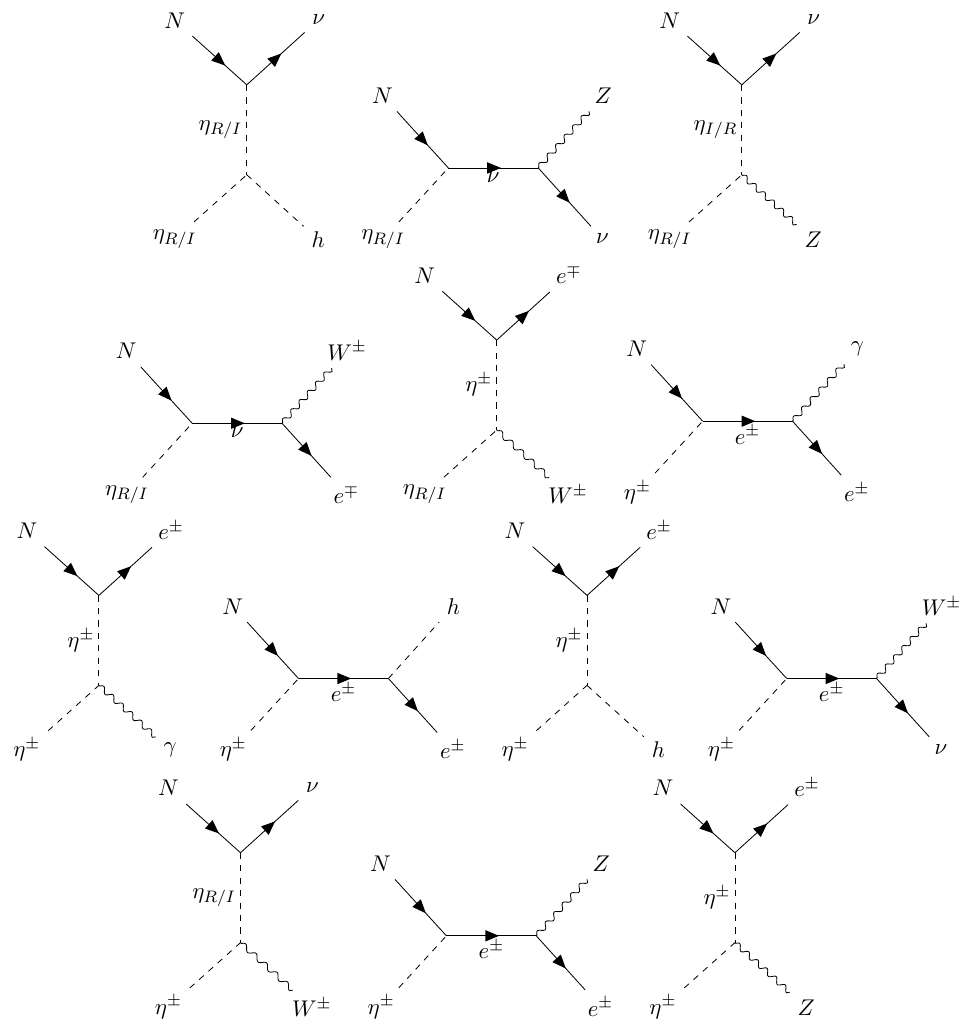}
		\caption{\centering Co-annihilation Channels}
		\label{fig:FermionCoAnnihilation}
	\end{subfigure}
	\caption{Annihilation (a) and co-annihilation (b) Feynman diagrams that contribute to the relic density of the fermionic dark matter candidate $ N $.}
\end{figure}

In Fig. \ref{fig:ScalarAnnihilation}, we show the annihilation and co-annihilation channels which determine the relic density of the scalar dark matter candidate $ \eta_R $. They include s-channel co-annihilation diagrams with $ \eta_I $ and $ \eta^\pm $ mediated by the gauge bosons into leptons and annihilation diagrams mediated by the Higgs boson into gauge bosons. There are also t-channel annihilations mediated by $ \eta^\pm, \eta_R $ and $ N $ into SM particles as well as annihilations into gauge bosons through quartic couplings.
In Fig.\ref{fig:direct}, we show the tree-level diagrams that contribute to the spin- independent $\eta_R$-nucleon elastic scattering cross-section mediated by the Higgs boson $ h $ and the $ Z $ boson.
In Fig. \ref{fig:FermionAnnihilation}, we show the annihilation and co-annihilation channels which determine the relic density of the fermionic dark matter candidate $ N $. They include s-channel co-annihilation diagrams with $ \eta_R, \eta_I $ and $ \eta^\pm $ mediated by leptons into SM particles. The t-channel co-annihilation diagrams are mediated by $ \eta^\pm, \eta_R $ and $ \eta_I $. The annihilation diagrams are all t-channel processes and they have relatively low cross-sections.

%

\newpage

\bibliographystyle{utphys}
\bibliography{bibliography} 
\end{document}